\RequirePackage{fix-cm}
\documentclass[smallextended]{svjour3}       % onecolumn (second format)
\smartqed  % flush right qed marks, e.g. at end of proof
\usepackage[round]{natbib}
\usepackage{psfig}
\usepackage{a4}
\usepackage[english]{babel}
\usepackage{graphicx}
\journalname{Foundation of Physics}

\newcommand{\solm}{M$_{\odot}$}

\begin{document}

\title{The Milky Way's Supermassive Black Hole:\\
How good a case is it?}
\subtitle{\it A Challenge for Astrophysics \& Philosophy of Science}

\titlerunning{The Galactic Center Black Hole}        % if too long for running head

\author{Andreas Eckart$^{a,b}$
Andreas H\"uttemann$^{c}$,
Claus Kiefer$^{d}$,
Silke Britzen$^{b}$,
Michal Zaja\v cek$^{a,b}$,
Claus L\"ammerzahl$^{e}$,
Manfred St\"ockler$^{f}$,
Monika Valencia-S.$^{a}$,
Vladimir Karas$^{g}$,
Macarena Garc\'{i}a-Mar\'{i}n$^{h}$
}

\authorrunning{Eckart et al.}

\institute{
Corresponding author: Andreas Eckart; eckart@ph1.uni-koeln.de\\
a) I. Physikalisches Institut, Universit\"at zu K\"oln, Z\"ulpicher Stra\ss e 77, 50937 K\"oln, Germany\\
b) Max-Planck-Institut f\"ur Radioastronomie, Auf dem H\"ugel 69, 53121 Bonn, Germany\\
c) Philosphisches Seminar, Albertus-Magnus-Platz 1, 50923 K\"oln, Germany\\
d) Institut f\"ur Theoretische Physik, Z\"ulpicher Stra\ss e 77, 50937 K\"oln, Germany\\
e) Center for Applied Space Technology and Microgravity (ZARM), Am Fallturm, 28359 Bremen, Germany\\
f) Universit\"at Bremen, Institut f\"ur Philosophie, FB 9, Postfach 330 440, Enrique-Schmidt-Str. 7, 28359 Bremen, Germany\\
g) Astronomical Institute, Academy of Sciences, Bo\u{c}n\'{i} II 1401, CZ-14131 Prague, Czech Republic\\
h) European Space Agency (ESA/STScI), 3700 San Martin Drive, Baltimore, MD 21218, USA
}

\date{Received: date / Accepted: date}
% The correct dates will be entered by the editor

\maketitle

\begin{abstract}

The compact and, with 4.3$\pm$0.3$\times$10$^6$\solm,
very massive object located at the center of the Milky Way is
currently the very best candidate for a supermassive black hole (SMBH)
in our immediate vicinity.
The strongest evidence for this is provided by measurements of
stellar orbits,  variable X-ray emission, and strongly
variable polarized near-infrared emission from the location of the radio source
Sagittarius~A* (SgrA*) in the middle of the central stellar cluster.
Simultaneous near-infrared and X-ray observations of
SgrA* have
revealed insights into the emission mechanisms responsible for the
powerful near-infrared and X-ray flares from within
a few tens to one hundred Schwarzschild radii of such a putative
SMBH at the center of the Milky Way.
If SgrA* is indeed a SMBH it will, in projection
onto the sky, have the largest
event horizon and will certainly be the first and most important
target of the Event Horizon Telescope (EHT)
Very Long Baseline Interferometry (VLBI) observations currently being prepared.
These observations in combination with the infrared interferometry experiment GRAVITY
at the Very Large Telescope Interferometer (VLTI) and other experiments across the
electromagnetic spectrum might yield proof for the presence of a
black hole at the center of the Milky Way.
The large body of evidence continues to discriminate the identification of SgrA*
as a SMBH from alternative possibilities.
It is, however, unclear when the ever mounting evidence for
SgrA* being associated with a SMBH will suffice as a convincing proof.
Additional compelling evidence may come from future gravitational wave observatories.
This manuscript reviews the observational facts, theoretical grounds
and conceptual aspects for the case of SgrA* being a black hole.
We treat theory and observations in the
framework of the philosophical discussions about 
``(Anti)Realism and Underdetermination",
as this line of arguments allows us to describe the situation in observational
astrophysics with respect to supermassive black holes.
Questions concerning the existence of supermassive black holes and in particular SgrA* are
discussed using causation as an indispensable element.
We show that the results of our investigation are convincingly mapped out by this combination of
concepts.
%First keyword \and Second keyword \and More}
% \PACS{PACS code1 \and PACS code2 \and more}
% \subclass{MSC code1 \and MSC code2 \and more}
\end{abstract}

\section{Introduction}
\label{Introduction}

The compact radio source Sagittarius~A* (SgrA*) is located within a dense
central stellar cluster surrounded by a small ($\sim$1 arcsecond in angular and 0.04 parsec linear diameter)
cluster of high velocity stars \citep{Reid2003, Ghez2008, gillessen2009a}.
SgrA* is active across the entire electromagnetic spectrum and subject to detailed investigations 
\citep[e.g.][]{genzel2010, Ghez2009, Eckartbook2005}.
In this contribution, we summarize  some of the
most intriguing characteristics of the central dark mass 
of 4.3$\pm$0.3$\times$10$^6$\solm
\footnote{In the context of the Galactic Center, the black hole mass is given in millions
of solar masses (\solm= 1.98$\times$10$^{30}$~kg).}
associated 
with SgrA* and the stars in its vicinity as well as the conceptual aspects 
- including a summary of its philosophical background -
that may lead to a convincing proof for the existence of a supermassive black hole (SMBH). 
Our knowledge on the Galactic
Center (GC) profits greatly from the very recent results obtained 
with large 8-10~m class ground based telescopes that operate in the infrared\footnote{e.g. the European Southern Observatory's 
Very Large Telescope on Paranal in Chile or the W.M. Keck Observatory's Keck telescopes on Mauna Kea in Hawai'i, USA}.
Only at these wavelengths it is possible to observe the high velocity stars 
and to monitor the optically thin synchrotron radiation from its immediate vicinity.
At infrared wavelengths, the dust and gas along the $\sim$8~kpc line of sight
towards the center can be penetrated, resulting in detailed data on the
stars and the infrared counterpart of SgrA*. 
Additional observational data stem from highest resolution radio-interferometric 
observations in the mm/sub-mm bands. 
A new and for this purpose specialized array of mm-antennas - the Event-Horizon-Telescope (EHT\footnote{The EHT is a Very Long Baseline Interferometry (VLBI) array 
working at millimeter wavelengths, and dedicated for
observing the event horizons (but see section \ref{subBHinGR}) of the largest SMBHs in the sky;
http://www.eventhorizontelescope.org/}) -
has already performed first observations of 
Sgr A* at 1mm and supplementary Very Long Baseline Interferometry (VLBI) 
information at 3mm and longer wavelengths is available.
Strong variability from the radio through the infrared to the X-ray domain also 
indicates that energetic processes like accretion onto a black hole may be dominating the 
center of the Milky Way.

Although the observational evidence for SgrA* being a SMBH
is ever increasing, one has to ask the question: 
{\it At which point do we have enough evidence to claim that SgrA* is a SMBH?}
This requires consensus on testable predictions based on a definition of a SMBH
as well as on the epistemological processes that need to be followed.
Therefore, the paper has two focal points: One on the philosophy of science
and the other on an analysis of current theoretical and observational facts that need
to be considered within the philosophical framework.

SMBHs are philosophically interesting entities given that they
are only observable by indirect means.
Within the philosophy of science, the question on the
reality of such unobservable entities, historically mostly
part of the microcosm, has a long tradition
\citep[see e.g. ][for a review and the Appendix~\ref{App:subPhilQuestions} for an example]{Chakravartty2015}.
We note that Ian Hacking has stressed
a special status for astrophysics in the discussion
of scientific realism \citep{Hacking1989}.
As a logical consequence of his experimental criteria for
realism \citep[see][]{Hacking1983,Sandell2010}
that require the possibility of interaction with and manipulation
of the entity in question, he is committed to astrophysical
anti-realism due to the observational nature of astrophysical research.
In particular, he uses the existence of not directly observable
gravitational microlenses in favor of his argument.
From the existence of gravitational microlenses he deduces an
inherent underdetermination of astrophysical theories,
that makes their results apparently less reliable in comparison with
experimental disciplines that allow for a direct interaction with 
the entities under investigation.

For the current paper it needs to be clarified
if supermassive black holes are
different entities compared to, for example electrons, chairs, or galaxies,
such that a special consideration in the framework of 
philosophy of science is justified.
While all of them can be discriminated from each other in the
framework of an constructive empiricism,
there are noteworthy differences:
Entities like chairs or tables stand out with respect to the
imperfection of their definition 
\citep[as outlined by, for example, David Hume; see][]{Nelson2010},
galaxies stand out with respect to their distance and complexity 
\citep[these object have been considered by][]{vanFraassen1985},
and electrons are very simple entities that 
can be ``sprayed" \citep{Hacking1989},
this means, they are readily available in the laboratory for causality test.
Black holes, and in particular supermassive black holes
stand out in the simplicity of their mathematical definition 
that is best expressed in the ``no-hair" theorem
(see section \ref{definition}).
They are not readily available in the laboratory and cannot 
easily be ``sprayed".
They are characterized by an event horizon that, however, cannot 
become part of an external observer's past in a finite time
but is an important discriminator against other
similarly compact and massive objects.

Claiming that black holes are only an artifact of classical theory 
and thereby
compromising on the seriousness with which we treat the event horizon
makes the problem even worse. In this case, it is even more difficult
to distinguish between the black hole solution and other descriptions of
compact objects of the same mass.

Black holes may, hence, be seen as the prototype of 
astrophysical entities that suffer from underdetermination.
There is, however, ample indirect evidence for their existence.
Still, as motivated by Hacking's line of arguments, it may be
interesting to raise the epistemological question of whether
we face a situation of underdetermination in the search for SMBHs.
How strong is the available evidence in favor of the
SMBH interpretation, being one possible ``causal story" to
unify the ``existing evidentiary traces" 
\citep[e.g. ][]{Cleland2002, Anderl2015, Anderl2016, Smeenk2005}.
Are there alternative theories that may as well serve as
possible explanations for our observations?
If yes, what kind of observations would we need to disprove
alternative theories or decide in favor of one of them?
How can we describe and organize the epistemological situation
we face in a formal way?

In order to investigate these questions,
we first summarize the theoretical predictions 
and initial observational evidences for black holes 
and, in particular, the SMBH at the center of the Milky Way in section~\ref{predictions}.
This includes the history of the evolution of the subject and considerations.
We then elaborate on the philosophical questions connected to the identification of SMBHs 
in section~\ref{PhilConcepts}.
In section~\ref{obsEvaluation}, we present a detailed description and evaluation of 
mostly observationally based results that point at the presence of a SMBH at the Galactic Center.
We then summarize possible alternatives to SMBHs in section~\ref{Alternatives} 
and mention their possible astrophysical consequences.
In section \ref{futureobs}, we summarize some of the important future possibilities 
that arise with new instrumentation.
A synthesis of all facts is attempted in section \ref{synthesis}.
In section \ref{conclusion}, we give a summary and conclusions in the framework of the 
philosophy of science principles outlined earlier.

\section{The concept of Black Holes}
\label{predictions}

Objects that can be so compact that light cannot escape their gravitational fields were first 
speculated on in the 18th century by John Michell and Pierre-Simon Laplace\footnote{For a detailed 
account for the history of black holes, see \cite{israel1987}.}.
In 1916, Karl Schwarzschild \citep{schwarzschild1916} 
found a solution of Einstein's field equations that
can describe a non-rotating black hole in vacuum, although not called like this at that time.
In the same year, Hans Reissner \citep{Reissner1916} found the solution  for a  
stationary charged point mass, which was extended by Gunnar  Nordstr\"om \citep{Nordstrom1918} 
for a spherically symmetric charged body.
In 1939, Einstein speculated that ultra-compact masses do not exist
\citep{Einstein1939}.
In 1939/40, the authors of \cite{OppenheimerVolkoff1939} and \cite{OppenheimeSnyder1939} worked on 
continued gravitational contraction and resulting objects.
It was David Finkelstein in 1958 \citep{finkelstein1958}
who highlighted that this was a region of space
from which virtually no information could escape.
These objects were referred to as ''frozen stars" at that time and
early relativists often used the term Schwarzschild's sphere, 
discontinuity, sph\'ere catastrophique, magic circle or just frontier or barrier \citep{Bartusiak2015}.
Soon after that it could be shown that black holes actually can generically be predicted from general relativity (GR).
Between the late 1940's and 1960's, Ann Ewing published articles on astronomy and 
physics in Science News \citep{brown2010}.
She used the term ``black hole" as early as 1964 in her article ``Black Holes in Space" \citep{ewing1964}
following a contribution at a meeting of the American Association for the 
Advancement of Science (AAAS) \citep{brown2010}.
The most recent investigation on the introduction of the term ``black hole" in astrophysics is given 
by \cite{Bartusiak2015} and summarized here briefly: 
She explains that, as reported by the editor of the Life Magazine Albert Rosenfeld, the astrophysicists
Fred Hoyle and William Fauler used the term at the 1963 Texas Symposium.
It was used again at the AAAS meeting a short time later by Hong-Yee Chiu from the Goddard institute.
Hong-Yee Chiu reports that the term may have been used around 1960 by 
the Princeton physicist Robert Dicke.
Ann Ewing had attended that very AAAS meeting in 1963.
This was quickly followed by the discovery of stellar black hole candidates.
\cite{Bartusiak2015} reports that John Wheeler may have been inspired by a poem 
titled ``Music of the Spheres" by A.M. Sullivan (1896-1980). The poem focuses on William Herschel and
uses the term ``black hole"\footnote{"When the long eye of Herschel Burrowed the heavens Near the 
belt of Orion He trembled in awe At the black hole of Chaos".}. It was published in the New York Times in August, 1967.
In that very year John Wheeler proposed to use the term ``black hole" as a technical term;
from that time on it became the standard term.

The theoretical structure of stationary black holes 
as solutions to Einstein's field equations was clarified in
the 1960s. Stationary black holes can be formed as the asymptotic end
state of a gravitational collapse. A most interesting result is the
validity of {\it uniqueness theorems} for black holes 
\citep[see e.g.][]{Heusler1996}.
Within the Einstein-Maxwell theory (describing
coupled gravitational and electromagnetic fields) one can prove that
black holes are uniquely characterized by only three parameters: mass
$M$, angular momentum $J\equiv cMa$ (see footnote~\footnote{For black holes the spin can be characterized by the angular
momentum parameter $a=J/Mc$, where
$J$ is the angular momentum, $M$ the mass of the black hole, and $c$ is the speed of light}
and details given below and in section \ref{Spin}), and electric charge $q$. In a
sense, black holes are thus much simpler objects than ordinary stars
-- stationary black holes with the same three parameters look the same. 
"Dark stars" that do not collapse can be shown to be unstable \citep{Cardoso2014}.
During the gravitational collapse of a star to a black hole, all
other degrees of freedom are radiated away, mostly in the form of
gravitational waves ("ringdown"). 
Since one can refer to these other degrees of 
freedom as ``hair", the uniqueness theorems are also known under the
name {\it no-hair theorems}.

The most general solution, with $M$, $a$, and $q$ non-vanishing, is
called Kerr-Newman solution. Since $q$ is thought to be irrelevant in
astrophysical situations, the most important case for astrophysics is
the Kerr solution, found by Roy Kerr in 1963 \citep{Kerr1963}; it is fully characterized by
$M$ and $a$. To the extent that the contribution of surrounding matter
is negligible, 
the black-hole candidates discussed in our paper are all
expected to be described by this solution. A black-hole solution with
$a=0$ is spherically symmetric and called Schwarzschild solution
($M\neq 0$, $q=0$) or Reissner-Nordstr\"om solution ($M\neq 0$, $q\neq 0$)
(see also comments in section \ref{charge}).

The concrete values of the few black hole parameters $M$, $a$, and $q$ 
must be obtained through observations of the black hole's interaction 
with its immediate environment. This is mostly accomplished by observing 
the strong electromagnetic radiation produced during these interactions.
The high luminosity of extragalactic sources found first at optical and then at radio wavelengths 
drew the attention to possible high mass concentrations in the nuclei of galaxies.
SMBHs were first introduced in order to explain the 
large luminosities of quasi-stellar objects (QSOs) and their radio counterparts called quasars.
These objects have been found to outshine their host galaxies at the centers of 
which they are located.

QSOs and quasars are believed to be powered by accretion of material 
into SMBHs \citep{ZeldivichNovikov1964, Salpeter1964, Rees1984}.
From first principles, it can be shown that this 
process very efficiently allows us to generate the required high luminosities.
Relativistic effects like superluminal motion (apparent motion faster than the
speed of light) in jets, black hole mass measurements
and high X-ray and $\gamma$-ray luminosities are further strong evidences for the presence of 
supermassive compact objects like SMBHs at the centers of galaxies.

In 1971, Donald Lynden-Bell and Martin Rees speculated on the presence of 
a SMBH at the center of the Milky Way. 
Three years later, the compact and stationary synchrotron radio source SgrA* was discovered
by \cite{BalickBrown1974} using radio interferometry.
Hence on theoretical and observational grounds, a SMBH
at the center of the Milky Way had to be seriously considered.
For the Galactic Center, the detection of gas streaming towards the center by
\cite{wollman1977} pointed at a high mass concentration.
The detection of high velocity stars presented first direct 
evidence for the presence of a SMBH
\citep{eckart1996nature, eckart1997, ghez1998}.

In the meantime, an impressive number of observations has yielded indirect evidence 
for the presence of black holes in the centers of many, if not all, bright galaxies 
\citep{ferrarese2006, nayakshin2012}.
Galaxy evolution itself seems to be strongly connected to black hole evolution 
\citep{ferrarese2006,Ferremateu2015,KormendeyHo2013}.
High masses of dark and very compact objects, i.e. most likely SMBHs 
(see section \ref{PhilConcepts} in particular section \ref{Alternatives}), 
can be measured, and it seems possible to determine the spin of the 
black holes from X-ray measurements \citep[e.g.][]{reynolds2014}.

Kiloparsec- and parsec-scale jets are among the most prominently visible signatures of black hole accretion. 
Cosmic black holes, including the SMBH in Sgr A*, are not isolated 
in their environment. 
Instead, surrounding gas and stars also contribute to some extent to the gravitational field. 
Therefore, the classical sequence of Schwarzschild - Kerr - Reissner-Nordstr\"om - Kerr-Newman 
solutions for black holes in vacuum should be generalized to account for distributed material 
in their vicinity. 
This has been discussed by many authors in the context of exact solutions in GR; 
nonetheless, self-gravitating disks and rings are of particular interest, as such 
structures are detected in many objects \citep{karas2004}.

Galaxy collisions and the subsequent coalescence of binary black holes explain galaxy/black 
hole growth and black hole activity \citep{sesana2011,bogdanovic2015}.
Strong gravitational wave signals occur during the final phase of the merging process 
\citep[see the detection of gravitational waves from binary stellar sources by the 
LIGO and Virgo collaborations][]{Abbott2016}
and might be detected with Pulsar Timing Arrays
\citep[e.g.][]{wyithe2003, roedig2012, pollney2011} in the future.
Current efforts are also summarized in reviews by
\cite{Berti2015},
\cite{Gair2013},
and \cite{Yunes2013}.
While the Pulsar Timing Array will detect the foreground 
of Gravitational Waves resulting
from the incoherent superposition of  inspiralling SMBH binaries
well before merging (i.e., far from coalescence),
LISA and eLISA\footnote{LISA and eLISA are proposed Laser Interferometer Space Antennae;
see https://www.elisascience.org/}
will detect SMBHs at coalescence \citep[e.g.][]{eLISA2013}.
These observatories will have
the highest sensitivity just in the mass range between 
10$^5$ to a few 10$^6$\solm, i.e., masses even below that of SgrA*.

Black holes play an increasingly important role in many areas of astrophysics.
Therefore, more pressing seems to be the question: How good a case is it?
With regard to collecting observational evidence closer and closer to the presumed 
event horizon it seems odd and intellectually dissatisfying to approach a zone of principal non-observability
(see also comments on ``underdetermination" in section \ref{subsection:antirealism}).
The wealth and complexity of the collected astrophysical observational
data on the surroundings and properties of the central object are in sharp contrast to the rather simple
expectations and predictions for the very central supermassive object. 
Yet, both regions are connected and need to form a consistent physical picture.
It is essential to explain in detail the tools of physics at the photon sphere of an event horizon,
and prepare for a consistent explanation of the wealth and 
complexity of the collected observational data 
describing the immediate vicinity of the central object.

\subsection{Working definition of a (supermassive) black hole}
\label{definition}

A black hole may be described as a geometrically defined compact region of space-time associated with a mass
\footnote{Recent aspects regarding quantum properties of very small black holes (which are not subject of our article but 
are important to get a complete picture of the 
envisaged properties of back holes) are given in e.g.
\cite{2016Christodoulou, Barrau2016, 2015Haggard, Rovelli2014, Barrau2014, Kiefer2015}.}.
The gravitation is so strong that nothing can escape from inside the event horizon
as defined in section \ref{subBHinGR}.
In this context, the no-hair theorem is often quoted which states
that a black hole is fully described by only three
externally observable classical parameters:
mass, electric charge, and angular momentum\footnote{
The same mathematical formalism allows us to include the magnetic charge (a monopole) 
as the ``fourth hair". However, this possibility is usually rejected on simple 
physical grounds - there are no magnetic monopoles found in nature so far.}.  
Of all known SMBH candidates in the Universe,
SgrA* at the center of the Milky Way is particularly well suited to
test the no-hair theorem, because it is the closest of these candidates
(see also comments on ``baryonic hair" in section \ref{macroquantumness}
and \cite{Johannsen2016}).  
A SMBH  differs from a stellar black hole simply by the
amount of mass contained in it. For stellar black holes the mass is 
typically between 1.4 and a few 10 solar masses, whereas the SMBHs 
located in the nuclei of massive galaxies harbor millions (to billions) of solar masses.
It is currently unclear how well the regime between the stellar and SMBHs
is populated. It is technically difficult to find very good candidates for these 
intermediate-mass black holes (IMBHs) in the mass range between
10$^2$\solm and  less than about 10$^5$\solm 
~~\citep{ColbertMushotzky1999}.
Ultra-luminous X-ray sources within the bulge and disks of galaxies 
or central objects in globular clusters are good candidates \citep[e.g.][]{Pasham2014, Worbel2016, Earnshaw2016, ArcaSedda2016}.
\cite{Kiziltan2017} found an indirect evidence for an IMBH of about 2200~\solm ~at the center of the globular cluster 47~Tucanae 
based on the dynamical  state constrained by the timing data of millisecond pulsars. 
However,  the inferred IMBH does not have a corresponding electromagnetic  counterpart, 
which implies that most of the IMBHs may be not accreting  enough to be bright in X-ray and 
radio domains, explaining the lack of  direct detections. 
In the Galactic plane, the IMBHs can also be  detected indirectly through either their velocity kicks imparted to the surrounding  molecular clouds
\citep{Oka2016}
 or the synchrotron signal associated with the bow shocks in case they  move supersonically \citep{Wang2014}.

The simplicity of the definition given above may be complicated by the fact that black holes may 
have an angular momentum (i.e. a spin) which leads to the formation of an ergosphere (from 
$\acute{\epsilon} \rho \gamma o \nu$,
 which means ``work" in Greek). 
In theory, it is possible to extract energy and mass from the ergosphere at the expense of the rotational 
energy of the black hole, which is also known as Penrose process or Blandford-Znajek effect for the 
explanation of the jet generation. The spin of the black hole also
determines the radius of the closest stable orbit of a (test) particle around it.
Within GR and the standard definition
of an isolated black hole, exact solutions of Einstein-Maxwell equations are known to 
describe black holes in terms of only the three above-mentioned physical parameters: 
mass, spin, and electric charge 
\citep{Chandrasekhar1983, DeWitt1973, HawkingIsrael1987}.
Astrophysically realistic solutions restrict also the mass of cosmic black holes 
by the mechanism and history of their formation. 
\cite{NatarajanTreister2009} show that there is an upper limit of 10$^{10}M_{\odot}$  
for astrophysical black holes. \cite{King2016}  continued in the analysis and derived a physical 
limit of $M_{{\rm  max}} \approx 5\times 10^{10}M_{\odot}$ to the mass of the  supermassive black hole, 
above which the luminous accretion cannot  proceed (the innermost stable circular orbit is larger than 
the  inner radius of the accretion disk). Hence, the maximum mass is just an observational limit,  
since further accretion above it is non-luminous, i.e. it does not manifest itself as a quasar or active galactic nucleus. 
Hence, black holes above $M_{{\rm max}}$ are expected to be low-luminous and therefore difficult to detect.

The allowed range of charge and spin is limited; 
above that limit, the horizon ceases and a naked singularity 
(see below in section \ref{subBHinGR})
emerges. 
In principle, a non-zero value of magnetic charge can be included; however, this 
possibility is usually neglected because of the (possible) non-existence 
of magnetic monopoles in the universe. 
When the black hole is embedded in a large-scale (cosmological) magnetic field, 
the standard assumption of asymptotic flatness must be abandoned and the notion 
of the event horizon has to be generalized 
\citep{GarciaDiaz1985, EstebanRamos1988}.
Nonetheless, other properties including the uniqueness theorems can be 
reformulated and maintained \citep{KarasVokrouhlicky1991}.

\subsection{Black Holes in General Relativity}
\label{subBHinGR}

The uniqueness theorems for black holes mentioned at the beginning of section~\ref{predictions}
refer to stationary black holes, that is, asymptotic end states of a collapse. 
For the general situation of non-stationary and non-symmetric black
holes, a more precise definition is needed. The exact mathematical
form is involved and can be found in standard textbooks such as 
 \cite{Wald1984}
or 
 \cite{Poisson2004};
for a semi-popular account, see \cite{Hawking1996}.
It captures the physical idea of a ``region of no escape", 
that is, a region of spacetime which will never become part of an
external observer's past. The boundary of this region is called {\em
  event horizon}, more precisely, {\it future event horizon}, because
one assumes the usual arrow of time (as given by the Second Law of
thermodynamics) to hold.\footnote{Because GR is
time-symmetric, there is also a past event horizon. It is the
boundary of a so-called ``white-hole region", that is, a spacetime region into
which nothing can enter and which thus will never become part of an
external observer's future. 
 White holes are usually excluded for
thermodynamic reasons, analogously to the exclusion of advanced
potentials in electrodynamics, cf.  \cite{Zeh2007} for a careful and
detailed discussion of these issues.} 

A future event horizon thus gives the boundary of the spacetime region 
(called black-hole region) from which nothing, not even light, can ever 
reach other spacetime regions. 
It must be emphasized that the whole future evolution must be 
known in order to specify the location of the event horizon;
the event horizon is a three-dimensional object (two spatial and one
lightlike dimension). 
But is the event horizon really the
most adequate concept for describing observations, as indicated, for
example, by the name of the project ``Event Horizon Project" (EHT; see introduction)?  
When observing a black hole
such as the SMBH in the Galactic center now, we cannot know of any
amount of matter that will fall into this black hole in the future and
will lead to an increase of mass and, consequently, of an increase of
the size of its event horizon. We thus need alternative notions which
are of a more local nature. Such notions are, in fact, 
used\footnote{Hawking has recently claimed that only apparent horizons exist \citep{Hawking2014}}.
The most important one for our case is the notion of an {\it apparent horizon}. 
For its definition, one considers the boundary between the region where 
emitted light can reach infinity and the region where it cannot. 
This three-dimensional boundary is called ``trapping horizon",
 and its two-dimensional intersection with a space-like surface 
(that is, with a surface of constant time), is called apparent horizon 
\citep{Wald1984, Poisson2004}.
This is the concept usually used in numerical relativity. 
The interior of the trapping horizon is characterized by the presence
of ``trapped surfaces"; these are two-dimensional surfaces for which
both in- {\it and} outgoing light rays propagate inwards\footnote{For the
observers of the black hole's ``shadow", it is perhaps the photon sphere that 
is the most important quantity.}. 
The word ``shadow" is used here to describe a region close to the black hole
towards which the emission is greatly suppressed due to relativistic effects 
(likelight bending and boosting due to light abberation).
A first formal approach as how to calculate photon paths near black holes 
in the context of a shadow was given by J.L. Synge in 1966 \citep{Synge1966}.
The first image of an accretion disk around a black hole including a 
shadow was given by Luminet in 1979 \citep{Luminet1979}.
Possible shapes and geometrical arrangements of the shadow are discussed, e.g., in
\cite{bardeen1973, Grenzebach2015, Abdolrahimi2015}.

For stationary
black holes of mass $M_{\bullet}$, the apparent horizon coincides with the (time slice of
the) event horizon. In the simplest case of the Schwarzschild
solution, the horizon size is given by the Schwarzschild radius
\begin{equation}
R_{\rm S}:=\frac{2GM_{\bullet}}{c^2};
\end{equation}  
for the Kerr black hole, the horizon is located at
\begin{equation}
R_{\rm Kerr}:=
\frac{GM_{\bullet}}{c^2}+\sqrt{\left(\frac{GM_{\bullet}}{c^2}\right)^2-a^2}.
\end{equation} 
Quite generally,\footnote{Provided the so-called ``null energy condition" holds} 
the apparent horizon lies {\it within} the
event horizon or coincides with it. 

The presence of trapped surfaces strongly indicates the formation of a
singularity, that is, a region where Einstein's equations (and 
other equations) no longer hold. This is the content of the famous
singularity theorems \citep{Hawking1996}.
Singularities are
predicted to occur in the interior of black holes. Strictly speaking,
their occurrence signals the need for a more fundamental theory that
replaces GR under such extreme conditions. It is
generally expected that this theory is a quantum theory of gravity,
see the next subsection.

It is an interesting and still to some extent open question whether an
event horizon forms in a realistic gravitational collapse. If not, the  
singularity (or its replacement in a more fundamental theory) will be
seen from outside, leading to a loss of predictability. Because of
the disturbing consequences of such ``naked singularities", 
Roger Penrose came up in 1969 with  his {\it cosmic censorship conjecture},
which loosely states that ``Nature abhors naked singularities" 
(see also comments on ``underdetermination" in section \ref{subsection:antirealism})
or, more precisely, in the formulation given in \cite{Wald1984} , p.~304: 
\begin{quote}
All physically reasonable spacetimes are globally hyperbolic, i.e.,
apart from a possible initial singularity (such as the ``big bang"
singularity) no singularity is ever ``visible" to any observer.
\end{quote}
Presently, it seems that this conjecture holds in a generic initial
condition (excluding conditions of high symmetry). If it were violated
and naked singularities did exist in Nature, this would have
tremendous consequences for astrophysics. The value of black holes (or
black-hole candidates) for the test of GR and possible
generalizations can hardly be overestimated 
 \citep{Yagi2016}.
This holds, in particular, for the Milky Way's SMBH, which can 
be viewed as an optimal ``laboratory" for GR 
 \citep{Johannsen2015}.

\subsection{Black Holes and Quantum Theory}
\label{subBHandQT}

We have seen that stationary black holes are fully characterized by
just three parameters. This reminds one of the situation in thermodynamics
where the macroscopic properties of a gas are fully characterized by
few parameters such as pressure, volume, and entropy. Surprisingly, it
was found in the 1970s that black holes {\it are} thermodynamic
systems; they obey complete analogies to all four laws of thermodynamics
(see e.g.
\cite{Hawking1996}
or 
 \cite{Kiefer1999}
for an
introduction). They possess, in particular, a temperature
(the {\it Hawking temperature}) and an
entropy (the {\it Bekenstein-Hawking entropy}), 
which can only be interpreted if quantum theory is taken into
account. In the case of a Schwarzschild black hole, the Hawking
temperature is given by
\begin{equation}
\label{TBH}
T_{\rm BH} =\frac{\hbar c^3}{8\pi k_{\rm B}GM_{\bullet}} 
\approx 6.17\times 10^{-8}
 \left(\frac{M_{\odot}}{M_{\bullet}}\right)\ {\rm K}\ .
\end{equation}
This leads to a finite lifetime for the black hole with an
order of magnitude estimate
\begin{equation}
\tau_{\rm BH}  \sim 10^{65}\left(\frac{M_0}{M_{\odot}}\right)^3\ {\rm years},
\end{equation}
where $M_0$ is the initial mass. The black hole becomes ``hot enough"
for quantum effects to be observable only near the final evaporation
phase, which for astrophysical black holes only occurs in a far
future. (For the Galactic Center black hole, the lifetime is of the order 
of $10^{84}$ years!) Nevertheless, there is also an astrophysical
interest in the study of Hawking evaporation. First, small black holes
may have formed in the early Universe and may evaporate today. Such
"primordial black holes" may even contribute to the dark matter in
the Universe 
\citep{Blais2002} and produce gravitational waves \citep{Bird2016}.
Second, according to recent
speculations, quantum effects may potentially become relevant near the
horizon of a macroscopic black hole; for example, in the form of a ``firewall"
 \citep{Calmet2015}; see also \cite{Giddings2014}.
It is an intriguing question whether such
effects, if present, could be observed with the EHT. 

Quantum effects of black holes play a crucial role in 
the search for a quantum theory of gravity \citep{Kiefer2012}.
Such a theory will give a complete description of black
hole evolution and, in particular, of the final evaporation phase. For
the latter, only very rudimentary models are available; see, for example,
 \cite{KieferMartoMoniz2009}.
The theory should also provide one with a
microscopic explanation of the Bekenstein-Hawking entropy. Whether
observations of the Galactic Center black hole can provide hints in this
direction, is an open issue.   
Let us finally emphasize that quantum effects are also essential 
for concepts of black hole alternatives (see section \ref{Alternatives}), 
like the ``fermion ball", ``boson star" , and ``grava-stars" scenarii,
as well as the phenomenon of macro-quantumness.

\section{Philosophical Concepts}
\label{PhilConcepts}
Although the observational evidence for SgrA* being a SMBH is ever increasing,
one has to ask the question: At what point can we be sure that it really is a SMBH?
Before we review the evidence in detail it is necessary to deal 
with some conceptual issues. 
First, what does it mean for a black hole to be real, or to exist? 
The philosophy of science allows us to use different approaches to deal 
with the problem of showing the reality or existence of SMBHs.
Here we follow a two stage approach:
First, we have chosen the concept of ``(Anti)Realism and Underdetermination" as one 
line of arguments that is close to physics. Observational astrophysics is a science of 
measuring and quantifying entities, which are subject to questions of 
(anti)realism and underdetermination
(see section \ref{subsection:antirealism}).
These are concepts that
predominantly help us to show the validity of theories.
Second, as we are interested in possibilities of probing the existence of SMBHs
we have chosen a line of arguments that requires the causation 
as an indispensable element if one wants to talk about the reality 
and/or existence of entities.
The linkage between these two lines or arguments is laid out in 
Appendix~\ref{App:section:bridge}.

\subsection{(Anti)Realism and Underdetermination}
\label{subsection:antirealism}

On one hand, philosophy of science has recognized the problem that the evidence 
from data may never be able to sufficiently discriminate one 
theory from another, rivaling theory. 
This problem of underdetermination is usually referred to as the 
epistemological problem of the indeterminacy of data to theory.
On the other hand, ``entity realism" is not committing itself to judge 
on the unique validity of scientific theories, but relies on repeatable 
routinely observable effects that provide or represent sufficient evidence to 
reliably support the validity of a theory.
This is reflected in an often mentioned quotation by Ian Hacking:
"if you can spray them, then they are real" \citep[][page 24]{Hacking1983}.
This statement can be translated into: If you can use entities to manipulate others,
then we have sufficient evidence for their reality.
Ian Hacking, as the main proponent of this formulation of entity realism, 
was referring to an experiment he observed in a
Stanford laboratory, where electrons and positrons were
sprayed onto a super-conducting metal sphere.
To be used as an instrument in a manipulation of other systems, this entity
realism presupposes a quantitative precise causal profile 
\citep[see e.g.][]{HoeferSmeenk2016} in order to bring 
about the effects in question.
If the effect is successfully brought about, then one has sufficient evidence 
for the claim that there is something with this particular profile.

Studying black holes as research objects faces a severe problem,
as global properties of a single, possibly uniquely defined object,
may be systematically underdetermined
\citep[e.g. ][]{Cleland2002, Anderl2015, Anderl2016, Smeenk2005}
That makes it difficult to find the “true” (responsible) causes for an 
observational result.
This underdetermination was identified by Hacking as a 
"methodological otherness" in comparison with other
realistic scientific disciplines
\citep[][]{Hacking1983,Sandell2010}.

Underdetermination has a theoretical and an experimental side: 
The theory may not be fully complete and only highlight certain properties.
The concept of underdetermination may indeed be particularly relevant 
for black holes as astrophysical objects,
for which - according to Hacking - we cannot make a claim to realism,
implying a sort of antirealism, as one may state that due to 
observational insufficiencies
we have no access to a reality even if it may exist.
As an example specific to black holes, the whole future evolution must be 
known in order to determine the location of the event horizon
(see also comments on ``naked singularities", and ``firewalls" in section
\ref{definition} and \ref{subBHinGR}, which will be similarly difficult to
observe and hence are likely to be subject to underdetermination).
Hence, at least in this case, underdetermination may be included as an inherent 
property of the theory.
In addition, the observations may be not unique enough to clearly 
distinguish one possible realization of an object from another,
since the interpretation of the observations 
may just be based on a restricted set of theoretical predictions.
In the case of experiments \citep[see e.g.][]{FranklinBook2016, Galison1987},
 however, one has the chance to fight 
(i.e. minimize or even remove the effect of) underdetermination by 
increasing the observational evidence and combining various procedures 
that approach the problem with different methods or instrumental efforts.

Currently, it is observationally quite possible to decide uniquely on 
the presence of very compact and massive objects based on astrophysical 
observations.
However, it is not quite possible to differentiate between different theories
of compact objects that predict similar observational facts
under the boundary condition of a minimal use of auxiliary input 
(i.e. theoretical stability arguments and rejection of the existence 
of certain particles; see section~\ref{Alternatives}).
This means that, despite theoretical and experimental progress,
it still remains difficult to resolve problems of  
underdetermination. 
This is reflected in the fact that the attempt of an observational 
proof of a theory involves and entails auxiliary hypotheses,
available instruments, and background assumptions 
\citep[e.g. ][]{LaudenLeplin1991}. 

An important feature that helps to minimize underdetermination is
the application of independent experimental observational methods or
channels. In this respect the newly established possibility to measure
gravitational wave signals is an essential point \citep[e.g.][]{Berti2016}.
While gravitational wave astrophysics is in its infancy, the possibility of
carrying out gravitational wave spectroscopy cannot be ignored when
discussing underdetermination. The cooperative efforts coming from
both ``classical", electromagnetic observational astrophysics
and the new gravitational wave astrophysics will substantially contribute
to our future picture of supermassive black holes.
However, \cite{Cardoso2016} point out that very compact objects with a 
light ring will display similar ringdown signals as black holes,
implying that universal ringdown waveforms indicate the presence of light rings, 
rather than that of horizons. 
Hence, it is only that high precision measurements of the late-time ringdown signals
may allow us to rule out exotic alternatives to black holes.

The situation depicted above is usually described in the framework 
of a ``contrastive underdetermination",
meaning that  for each set of observational evidences 
there may exist several different but equally empirically testable theories.
Stanford (2013) points out that this ``holist underdetermination" implies 
that a theory can be adopted to new sets of observational results.
This is particularly true, as these results do not solely depend
only on observational facts that test the theoretical claims, but also depend 
on many other auxiliary hypotheses, which may be right, or wrong or which 
may be right by themselves but cannot be fully used in the context of a test.

In the case of supermassive black holes this may lead to
an ``abductive justification" for black hole proofs,
i.e., the assumption that a theory might very likely be true
if it is the currently best explanation of the 
observational facts (including a set of auxiliary information).
This asks for a strategy that tries to reduce the holistic underdetermination 
by investigating different sets of auxiliary assumptions for their influence 
on the theoretical prediction and the interpretation of observational evidences.
This is our main intention in sections \ref{obsEvaluation}, \ref{Alternatives}, and
\ref{synthesis}.

\subsection{The Involvement of Causality}
\label{subsection:bridge}

The discussion on ``Realism" and ``Underdetermination" targets mainly the 
consistency and repeatability of observations and in how far they provide 
sufficient evidence to reliably support the validity of a theory.
As in all experiments discussed below it is never the supermassive black holes
themselves that are observed 
(with the gravitational radiation originating closest to the black holes;
see section~\ref{ringing})
but only objects that interact with it, e.g. gas, stars, light.
Hence, the evidence to reliably support the validity of a theory comes in through
causality described by the theory.
Therefore, for both theory and observations there are direct links from the concepts 
of ``Realism" and ``Underdetermination" to the concept of the causality.
More detailed information on how this linkage may be described is given in Appendix~\ref{App:section:bridge}.
Hence, if the problem of underdetermination can be minimized, one is approaching the
situation in which the causal relevance and the indispensability of black holes 
is strengthened.  
More detailed information on this is given in Appendix~\ref{App:section:EleaticPrinciple}.
in which we describe the Eleatic Principle as a possible causal criterion that may be used
\citep[][]{Colyvan1998}.

\begin{figure}
\begin{center}
\includegraphics[width=12cm]{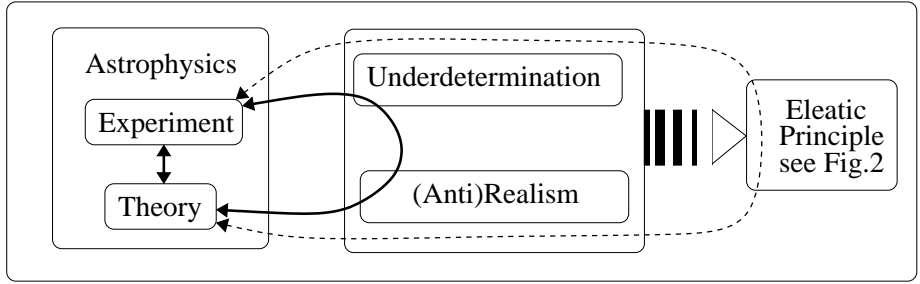}
\end{center}
\caption{
Linkage between experiment and theory interpreted via the concept of
realism and underdetermination finally allowing us to discuss the  question of
realism and existence in the framework of causation, making use of a form
of the Eleatic Principle.
\label{figm00}} 
\end{figure}

The overall concept is shown in Fig.\ref{figm00}, showing the linkage 
(the thin continuous black lines) 
between experiment and theory that can be discussed using both, the concept of
realism and underdetermination.
If underdetermination can be minimized, we may approach 
(as shown by the dashed thick bar) 
a situation in which we can start discussing 
(shown by the thin dashed black curved line)
the question of underdetermination, realism and existence in the framework of causation,
justifying the usage the ``rounded out" version of the Eleatic Principle
(see Appendix~\ref{App:section:EleaticPrinciple}).
Usually a causal criterion is implemented as
a test that must be passed by logical statements or objects in order 
to be accepted within a
scientific context, that is, the study of the nature of being, 
becoming existent, existence, or reality.

\begin{figure}
\begin{center}
\includegraphics[width=12cm]{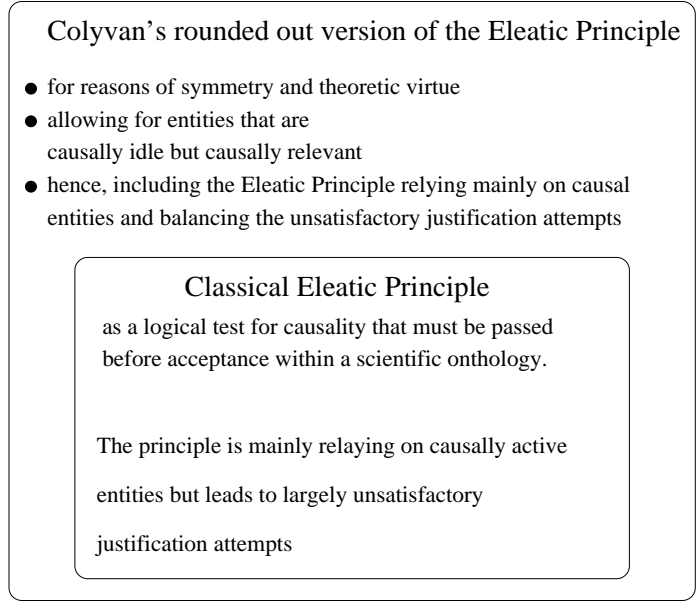}
\end{center}
\caption{
The classical Eleatic Principle in relation to the 
"rounded out" version proposed by
\cite{Colyvan1998}.
\label{fig00}} 
\end{figure}

\begin{figure*}
\begin{center}
\includegraphics[width=12cm]{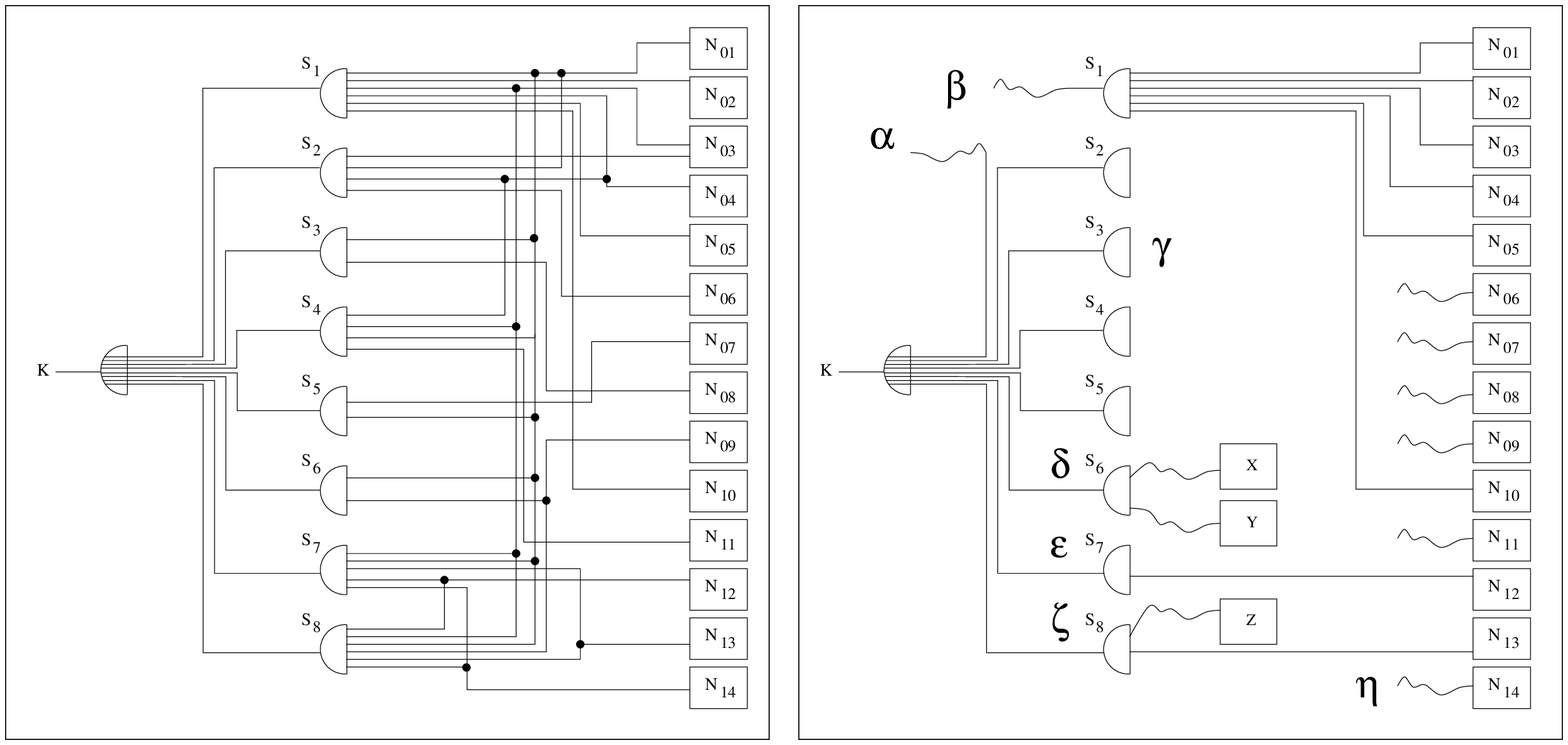}
\end{center}
\caption{{\bf Left:}
Epistemological process following the Eleatic Principle requesting logical standards of clarity as criteria of truth -
here represented by the usage of logical gates.  
In section~\ref{synthesis}, this network is used and described in detail.
Here appropriate necessary conditions $N_i$ are combined to sufficient conditions $S_k$ and a combined conclusion $K$ 
following logical standards (Here $i$ and $k$ are running indices).
{\bf Right:}
Depiction of how criteria simply based on sensual experiences may effect the epistemological process.
In the case of astrophysical research, we may consider data, i.e. observational or computational results, 
as sensual experiences.
If we do not follow causal criteria, then the usage of these data as sensual experiences does not 
stringently require the usage of logical standards to link them. 
We labeled a few cases that demonstrate what could go wrong:
$\alpha$ : the claim of existence is a claim not supported by data;
$\beta$ : a combination of correct necessary conclusions is (accidently) not considered;
$\gamma$ : a sufficient condition is claimed and not supported as reliable data;
$\delta$ :  a sufficient condition is  wrongly concluded from unrelated conditions;
$\epsilon$ : a sufficient condition is  (accidently) claimed from an insufficient number of necessary conditions;
$\zeta$ : a sufficient condition is  (accidently) claimed from a combination of unrelated conditions and related necessary conditions;
$\eta$ : properly related necessary conditions are overseen and not used.
\label{fig0}} 
\end{figure*}

While there is some debate about whether the causal criterion
provides a necessary condition 
for existence or reality (e.g. the reality of numbers in mathematics is disputable), 
there is much less controversy that it provides a 
sufficient condition for existence or reality. 
There is no question that there is something at the center of the Milky Way 
in virtue of there being something acting on the nearby stars etc.~.
The essential question is whether this something is a black hole. 
Under what conditions can we decisively answer this question?
For comparison we give in Appendix~\ref{App:subPhilQuestions} a historic example 
for establishing existence claims in the case of atoms and molecules.

Hence, in order to identify a specific entity, it is not sufficient 
to know that there is something that has some sort of effect. In order to establish a specific existence claim 
we need to establish that the entity in question manifests itself in very specific ways, so as for instance 
to put constraints on the values of experimentally accessible variables. 
An essential presupposition for establishing a specific existence claim is thus to have a sufficiently precise 
characterization of the entity in question (as done in section \ref{definition}), 
a characterization in terms of the causal/nomological relations it enters into. 
In order to get clarity about measuring and combining these relations we 
follow a formal logical 
approach outlined in the next section.

\subsection{Formal Logical Approach}
\label{formal}

There is a broad spectrum of different observational entities that either result from 
different instrumental methods and/or refer to different aspects of theory.
Naturally this results in a similarly broad spectrum of combinations of these entities 
to (observationally and theoretically) support the validity of the black hole theory and lead to the acceptance 
of the existence of supermassive black holes which is at the focus of this contribution.
To provide clarity in the approach we need to combine these entities in a 
formal logical way.
Such a treatment then fully accounts for the logical standards following the a causal criterion 
(e.g. the Eleatic Principle) opposed to criteria simply based on sensual experiences as 
demonstrated in Fig.~\ref{fig0}.
Probing the existence of entity $K$ in such a formal way, $K$ can be expressed 
via  the concept of necessary and sufficient conditions for something being fulfilled or true - within 
the philosophical concepts laid out in the previous sections.
In the following, we have to assume that $\nu$ and $\mu$ are natural numbers.
If $N_{\mu}$ are {\it necessary conditions} and if we use the universal quantifier,
we can then express the logical value of $S$ via:

\begin{equation}
S \Longrightarrow \forall \mu N_{\mu} \Longleftrightarrow  N_{1} \wedge N_{2} \wedge N_{3} \wedge ... \wedge N_{\mu}~~. 
\label{eq:1}
\end{equation}

The question may remain if all necessary conditions for
the proof of existence are known and fulfilled and if  the 
result then is indeed a sufficient condition for the existence.

Similarly if $S_{\nu}$ are {\it sufficient conditions} we can use the existential quantifier
and can then express the logical value of $K$ via:

\begin{equation}
K \Longrightarrow  \exists \nu S_{\nu} \Longleftrightarrow  S_{1} \vee S_{2} \vee S_{3} \vee ... \vee S_{\nu}~~. 
\label{eq:2}
\end{equation}

Here, the question remains if there are several sufficient conditions and if they are all of equal
quality, i.e., is one better or more convincing than the other to prove the existence
even though they both can in principle be true (or false).
There may also be the case that a sufficient condition may be 
represented via the concatenation of a subset of 
two or more necessary conditions, such as:

\begin{equation}
S_{i} \Longrightarrow \forall \lambda N_{\lambda}
\label{eq:3}
\end{equation}

\noindent
with $1\le i \le \nu$ and $1\le \lambda \le \mu$.
This allows us now to formulate a formal logic description 
of how to combine the necessary conditions
in order to test if SgrA* can be accepted as a SMBH.
In general, $\nu$ different subsets of 
a number of $\mu(\nu)$ 
necessary conditions may lead 
to sufficient conditions
with the logical value of $K$ expressed via:
\begin{equation}
K \Longrightarrow  \exists \nu S_{\nu}
  \Longleftrightarrow  \exists \nu \forall \mu(\nu) N_{\nu,\mu(\nu)}
\label{eq:4}
\end{equation}

\noindent
 with 

\begin{equation}
\begin{array}{lcl} \exists \nu \forall \mu(\nu) N_{\nu,\mu(\nu)} & \Longleftrightarrow &
(N_{\kappa_{1,1}} \wedge ... \wedge N_{\kappa_{1,\mu(1)}})_1 \vee \\
& & (N_{\kappa_{2,1}} \wedge ... \wedge N_{\kappa_{2,\mu(1)}})_2 \vee ... \\
& & \vee (N_{\kappa_{\nu,1}} \wedge ... \wedge N_{\kappa_{\nu,\mu(\nu)}})_{\nu}~~.
\end{array}
\label{eq:5}
\end{equation}

Here, $\nu$ is the number of combinations of necessary conditions to form a sufficient condition
while $\mu(max)$ is the total number of necessary condition available.
Then $\mu(\nu)$ is the number of necessary conditions for each of these  $\nu$ possible combinations
with  $\mu(\nu)$$\le$$\mu(max)$.
Then sufficient condition $S_{i}$ ($1\le i \le \nu$) is a combination of necessary conditions
with $N_{\kappa_{\lambda,\mu(\lambda)}}$ being 
the $\kappa_{\lambda,\mu(\lambda)}$-th element of the set of all necessary conditions for the $\nu$-th combination.

If for all combinations (i.e. for all $\lambda$ with $1\le \lambda \le \mu$)
e.g. one common necessary condition $C = N_{\kappa_{\lambda,\mu(\lambda)}}$  
 could be declared as a common necessary condition, this could of course be isolated, 
\begin{equation}
\begin{array}{lcl} 
K& \Longrightarrow & \exists \nu \forall \mu(\nu) N_{\nu,\mu(\nu)} \\
 & \Longleftrightarrow & C \wedge [ (N_{\kappa_{1,1}} \wedge ... \wedge N_{\kappa_{1,\mu(1)}})_1 \vee \\
 &                     & (N_{\kappa_{2,1}} \wedge ... \wedge N_{\kappa_{2,\mu(1)}})_2 \vee ... \vee \\
 &                     & (N_{\kappa_{\nu,1}} \wedge ... \wedge N_{\kappa_{\nu,\mu(\nu)}})_{\nu}]~~.
\end{array}
\label{eq:6}
\end{equation}
However, for simplicity it is easier to treat the complete sets
of necessary conditions, even though a more compact notification may be possible.
In section \ref{synthesis}, we will give examples for different combinations for the case of 
the SMBH at the center of the Milky Way.

\section{Evaluation of observationally based results}
\label{obsEvaluation}

\begin{figure*}
\begin{center}
\includegraphics[width=12cm]{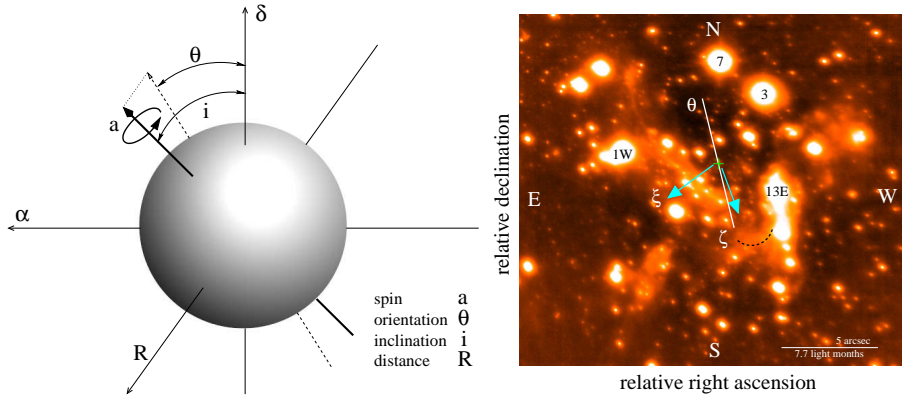}
\end{center}
\caption{{\bf Left:} Relation between the observational quantities spin, orientation, inclination and distance of the 
black hole. The sphere depicts the Schwarzschild horizon of the black hole.
The declination and right ascension axes are indicated by the symbols $\delta$ and $\alpha$, respectively.
The dashed line represents the projection of the spin axis onto the plain of the sky.
{\bf Right:}
An image of the central stellar cluster at 3.8$\mu$m wavelength (taken by some of the authors 
with the NACO instrument at the European Southern Observatory's Very Large Telescope) 
as it is surrounding the position of the SMBH SgrA* indicated by a  green cross.
The angular scale and the corresponding length scale at the location of the Galactic Center is given in the
lower right corner of the image.
Some bright stars (IRS1W, IRS3, IRS7 and IRS13E) are labeled.
Possible geometrical orientation of the SgrA* system derived from near-infrared 
polarization measurements by \cite{2015Shahzamanian} is indicated by the white line labeled
with the letter $\theta$.
The blue arrow labeled with the letter $\xi$ indicates the orientation of a putative jet.
The blue arrow labeled with the letter $\zeta$ indicates the orientation of a putative wind from
SgrA* into the mini-cavity shown as a dark dashed line.
\label{fig1}} 
\end{figure*}

In the following, we give a description and evaluation of observationally based 
facts that either point at the existence of a SMBH 
or are interpreted in the context of such an object.
Those include the 
distance,
mass,
size,
spin,
orientation,
spectrum,
and variability, 
as well as the possibility of a jet,
horizon,
shadow,
or a charge, to be described in sections \ref{Distance} to \ref{charge}.
On the left side of Fig.\ref{fig1} we show the relation between some of these quantities.
The right side of Fig.\ref{fig1} shows an infrared image of the central stellar cluster at the center of which SgrA* is located.

\subsection{Distance}
\label{Distance}

The distance to the center of the Milky Way and hence to the location of the SMBH 
is essential, for it allows us to transform the observed angular measurements into linear 
scales.\footnote{Angular scales in this context are
mostly given in fractions of arcseconds, i.e., milliarcseconds (mas) or microarcseconds ($\mu$as). While
the physical distance to the Galactic Center is given in kilo-parsecs 
(kpc; one parallactic seconds or 1~pc = 3.09$\times 10^{16}$~m) the physical scales 
at the Center are given in light years (ly = 9.46$\times$10$^{15}$~m) or parsecs 
(parallactic seconds or 1~pc = 3.3 ly) or fractions thereof, i.e.,
milli-parsecs (mpc) of micro-parsecs ($\mu$pc).
On even smaller scales, the sizes are given in astronomical units (AU = 149,597,871~km)
or in Schwarzschild radii $R_s$ that are linked to the black hole mass $M$ via $R_s$=$\frac{2GM}{c^2}$,
where $G$ is the gravitational constant and $c$ is the speed of light.}
It position in the sky has been determined very accurately via VLBI measurements
\citep{Menten1997, Reid2003}.
\footnote{Values from \cite{Menten1997}
with corrections by \cite{Reid2003}:
(J2000)
%R.A.=$17^h45^m40.098^s\pm0.03^s$, 
%Dec.=$-29^o00'27.08''\pm0.05$
}
There are two often used methods to derive the distance to the black hole at the center of the Milky Way.
The classical method uses globular clusters - or in general - stellar populations, 
determines their distances or kinematic 
properties and then infers symmetry considerations to derive the center of the Milky Way. Under the assumption that the 
SMBH is located at this center, it then can be taken as the distance to the black hole radio counterpart SgrA*.
The second method uses orbits of a single star (mostly S2) or several central stars and combines proper motions and 
radial velocity information to derive the distance of these stars orbiting the SMBH.

\begin{figure*}
\begin{center}
\includegraphics[width=12cm]{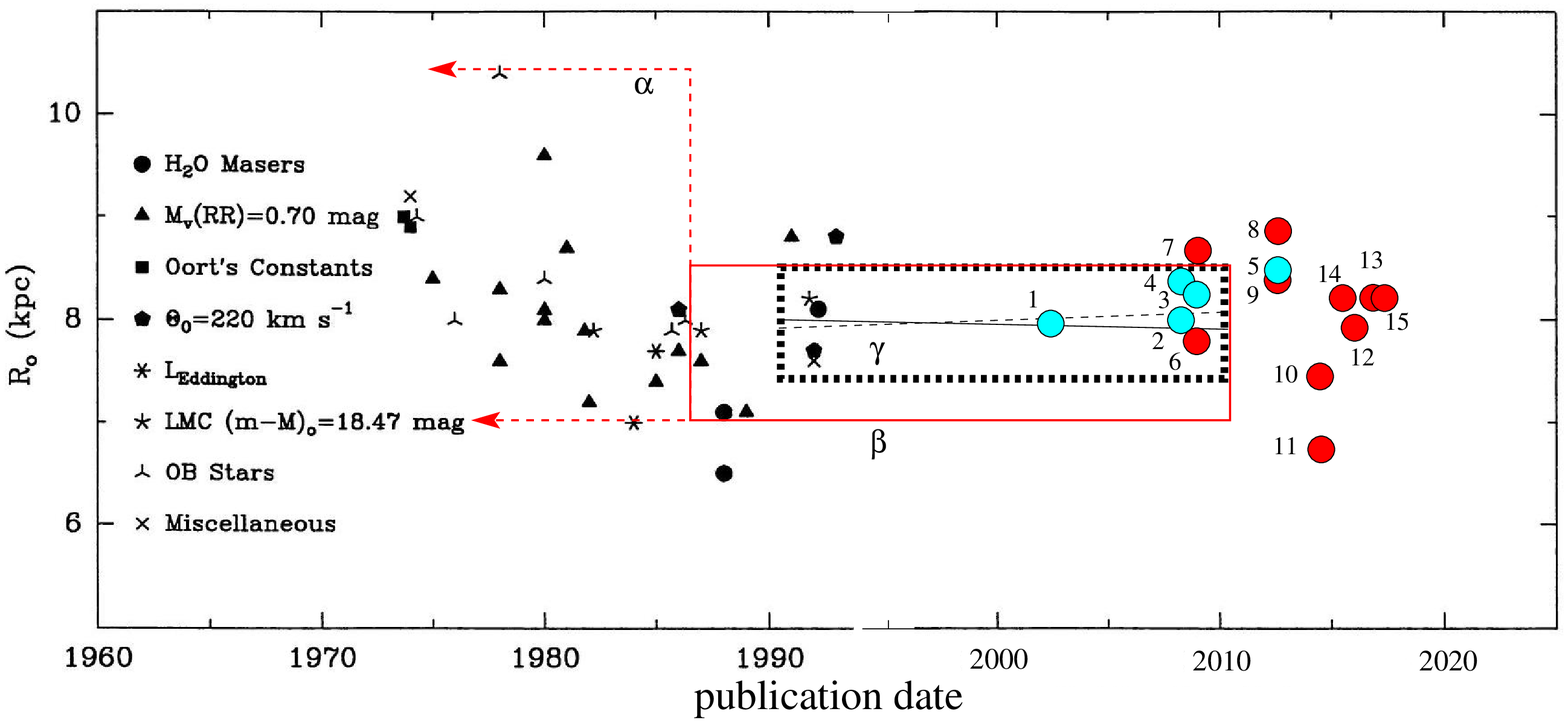}
\end{center}
\caption{Distance to the Galactic Center:
Black symbols and comments are taken from \cite{reid1993}.
The bulk of the data presented by
\cite{genzel2010} is located in region $\alpha$ and $\beta$.
The data used by \cite{Malkin2013} is mostly located in region $\gamma$.
The solid and dashed lines in region $\gamma$ correspond to the weighted and
unweighted versions of the calculated trends found by \cite{Malkin2013}.
New and recent data points based on stellar orbital analyses are 
shown by turquoise filled circles and labeled:
1: \cite{horrobin2004, eisenhauer2003, schoedel2002}; %  7.94+-0.42 S2
2: \cite{Ghez2008};                                   %  8.0 +-0.6  S2
3: \cite{gillessen2009a};                              %  8.28+-0.32 S2
4: \cite{Ghez2008};                                   %  8.46+-0.40 S2
5: \cite{do2013}.                                     %  8.46+-0.40 S2
Distance estimates based on stellar distributions in the global stellar cluster or Milky Way bulge are
shown by red filled circles and labeled:
6  \cite{Majaess2009};                                %  7.7 +-0.7
7: \cite{Vanhollebeke2009};                           %  8.7 +-0.71 
8: \cite{do2013};                                     %  8.92+-0.56
9: \cite{dekany2013};                                 %  8.33+-0.15
10:\cite{Francis2014};                                %  7.4 +-0.28
11:\cite{Branham2014};                                %  6.72+-0.39
12:\cite{Boehle2016};                                %  7.9+-0.2
13:\cite{Parsa2017}.                                %  8.25+-0.02
\label{fig2}} 
\end{figure*}

\cite{genzel2010}
compiled all distance estimates to the Galactic Center from the beginning of the last century until 2010.
\cite{Malkin2013} carried out a thorough and detailed study of the Galactic Center distance measurements between 1990 and 2010.
Both studies seem to suggest that the uncertainty in the distance estimate apparently decreases as a function of time.
In Fig.\ref{fig2} we summarize the recent  efforts to determine the distance to SgrA*.

From symmetry considerations of the 
distribution of recently updated globular cluster distances, \cite{Francis2014}
estimate the distance to the Galactic Center as R=7.4$\pm$0.28 kpc.
However, typically larger values are obtained if more information is 
included in the distance derivations.
From the distribution of several thousand OB-stars within 1~kpc of the Galactic Center, \cite{Branham2014}
derived a distance to it as R=6.72$\pm$0.39 kpc.
Using the characteristics of the stellar populations and the metallicity 
distribution in the Galactic bulge, \cite{Vanhollebeke2009}
derived a distance to the Galactic Center of R=8.7$\pm$0.71 kpc.
\cite{Majaess2009} use the distribution of $\delta$-Cepheids and find R=7.7$\pm$0.7 kpc.
\cite{dekany2013} combined optical and near-infrared data of known RR Lyrae (RRL) stars in the bulge and
studied the spatial distribution and distances of 7663 RRL stars. 
The authors obtained a distance to the Galactic Center of R = 8.33$\pm$0.15 kpc. 
\cite{dekany2013} also  report a different spatial distribution of the metal-rich and metal-poor stellar 
populations suggesting that the Milky Way may have a composite bulge. 
Depending on the nature of this bimodal distribution, this may also indicate that more indirect methods of deriving the distance to the 
Galactic Center may be affected by systematic uncertainties.
For the central 0.5 pc of the Milky Way nuclear star cluster  
\cite{do2013} present three-dimensional kinematic observations of stars 
using adaptive optics imaging and spectroscopy from the W.M.Keck telescopes. 
Solving simultaneously for the de-projected spatial density profile, cluster velocity anisotropy, 
the black hole mass, and distance to the Galactic Center they find 
a value of R=$8.92\pm0.56$ kpc.
In combination with the orbital data of S2 an improved value of 
R=$8.46\pm0.40$ kpc is obtained.

In the mean time, extensive astrometric and spectroscopic observations of the central
high velocity stars have been carried out
using the  ESO 8.2~m mirror Very Large Telescope dishes and the W.M.Keck~10~m telescopes
\citep{schoedel2002, eisenhauer2003, horrobin2004, Ghez2008, gillessen2009a, gillessen2009b}.
The most recent quite consistent combined estimates of mass and distance have been obtained by
\cite{Boehle2016} with 
$M = 4.0\pm0.2 \times10^6~M_{\odot}$ and $7.9\pm0.2$~kpc based on 
a combined orbital fit of S2 and S38 data,
and by \cite{Parsa2017} with
$M = 4.29\pm0.02\times10^6~M_{\odot}$ and $8.25\pm0.02$ kpc based on 
a combined orbital fit of three stars closest to SgrA*: S2, S0-102 and S38.
From Fig.\ref{fig2}, we see that the measurements (as a function of time) started off from higher values of
up to 10~kpc and have now settled to a value of about 8~kpc with total uncertainties of the order of 0.2~kpc.

\subsection{Mass}
\label{Mass}
The mass
is one of the fundamental quantities that determine the nature of a black hole.
Gaseous and stellar probes can be used to determine this quantity. 
Through observations of the 12.8$\mu$m NeII line emission from the mini-spiral in the Galactic Center 
stellar cluster \cite{wollman1977} revealed an enclosed mass of 4$\times 10^6$\solm.
However, at this time it was unclear how much the stellar cluster would contribute to the enclosed mass.
These measurements were also conducted using ionized gas as mass probes. Hence pressure gradients and 
magnetic field could have influenced the result in addition. 
Compact point masses are much better probes for a gravitational potential, i.e., a mass measurement.

\cite{krabbe1995}
report the first results of an extensive new study of stellar radial velocity measurements of the Galactic Center stellar cluster.
The authors have measured the radial velocity dispersion of 35 stars with distances of less than 12`` from 
Sgr A* as $154\pm19$~km~s$^{-1}$. 
This results in a mass estimate of about $3\times 10^6$\solm within 0.14 pc of the dynamic center.
The first stellar proper motion measurements were reported by
\cite{eckart1996nature}, \cite{eckart1997}, and \cite{ghez1998}.
Here, the authors present proper motion of 39 stars located between 0.03 and 0.3 pc from the Galactic Center.
These motions indicate the presence of a central dark mass of $2.45\pm0.4 \times 10^6$\solm ~~located within 0.015 pc
of the compact radio source Sgr A*.
This compares well with the mass of  $2.45\pm0.4 \times 10^6$\solm ~~derived by \cite{ghez1998} from proper motions within the
same region.
Combining the first stellar orbital acceleration measurements from W.M.Keck data
\citep{ghez2002} and from SHARP/NTT data \cite{eckart2002} apply a first order projection correction and 
derive a mass of M$_{acc} =(5 \pm 3) \times 10^6$\solm.
This estimate is consistent with the enclosed mass range of (2.6-3.3)$\times 10^6$\solm ~~obtained by
\cite{genzel2000} from radial and/or proper motion velocities of a homogenized sample of sources. 
Combining NTT and VLT data for the S2 orbit 
several authors \citep{schoedel2002, horrobin2004, eisenhauer2003} find
M=$3.7 \pm 1.5 \times 10^6$\solm. 

From three-dimensional stellar data within the central 0.5 pc of the Milky Way, \cite{do2013}
derive a SgrA* mass of M=$5.76_{-1.26}^{+1.76}\times 10^6$\solm.
Based on more than a decade of astrometric measurements of stellar orbits in the central half parsec,
\cite{Ghez2008} find a mass of M=4.5$\pm$0.4 \solm under the assumption that the SMBH 
is at rest with respect to the stellar cluster.
\cite{gillessen2009a} derives from the combined W.M.Keck and VLT data set a black hole mass of 
M=$4.30 \pm 0.50 \times 10^6$\solm.

\begin{figure*}
\begin{center}
\includegraphics[width=12cm]{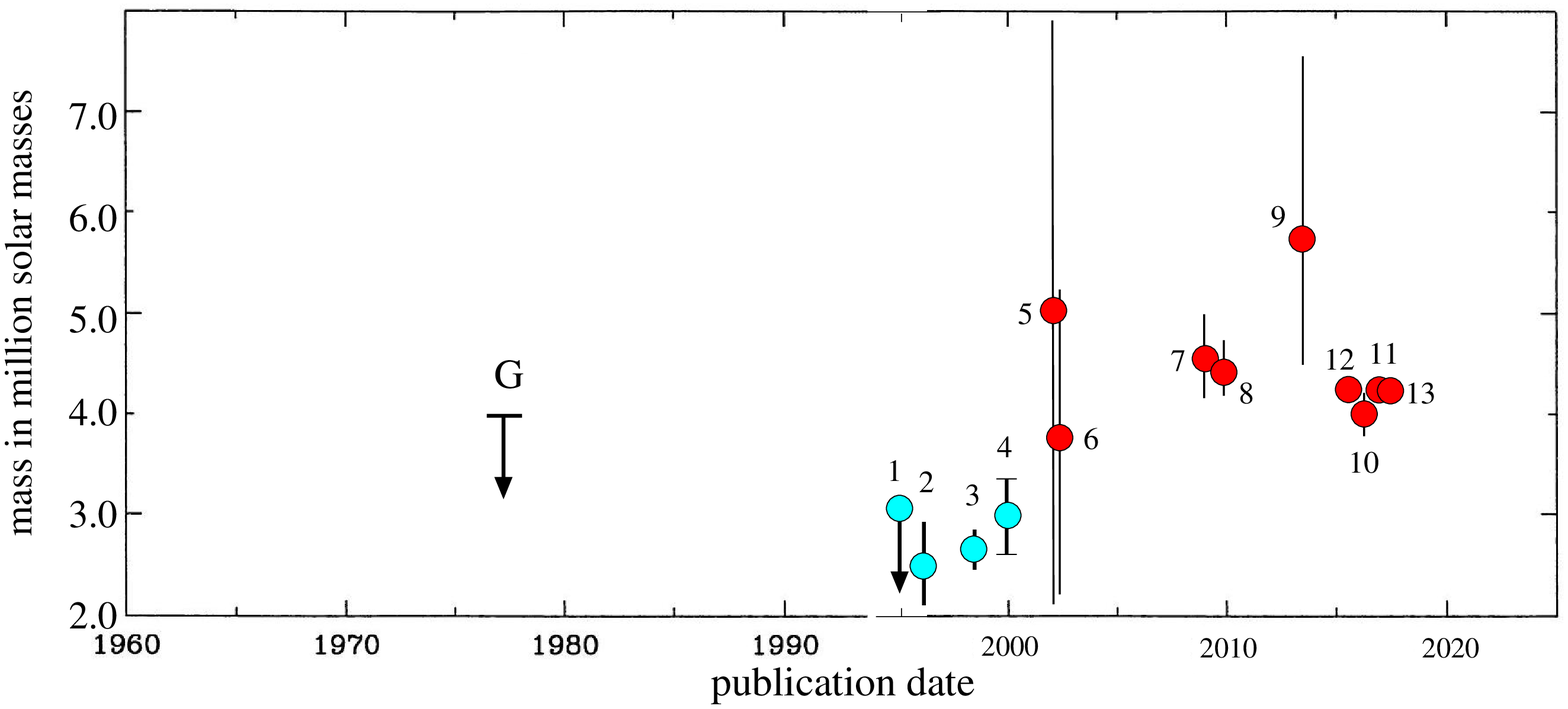}
\end{center}
\caption{Mass of SgrA* at the Galactic Center:
The enclosed mass estimate derived from mini-spiral gas measurements by 
\cite{wollman1977} is marked with a bold face ``G".
Mass estimates derived from stellar radial velocity and proper motion measurements are 
shown by  turquoise filled circles and labeled:
1: \cite{krabbe1995};                              % 3
2: \cite{eckart1996nature};                        % \cite{eckart1997}  2.45 0.4
3: \cite{ghez1998};                                %  2.6 0.2
4: \cite{genzel2000}.                              % (2.6-3.3)$\times 10^6$\solm
Mass estimates derived from orbital curvatures and stellar orbits are 
shown by red filled circles and labeled:
5: \cite{eckart2002};                              % 5+-3 curvature
6: \cite{schoedel2002, horrobin2004, eisenhauer2003};% 3.7  +-1.5
7: \cite{Ghez2008};                                % 4.5$\pm$0.4 
8: \cite{gillessen2009a};                           %  4.30 \pm 0.20 \pm 0.30 
9: \cite{do2013};                                  %5.76_{-1.26}^{+1.76}
10:\cite{Boehle2016};                                %  4.0+-0.2
11:\cite{Parsa2017}.                                %  4.29+-0.02
\label{fig3}}
\end{figure*}

In Fig.\ref{fig3}, we summarize the  efforts to determine the mass of SgrA*.
Historically, the problem was to disentangle determinations of the black hole mass from potential 
contributions of the stellar cluster surrounding it. 
The graph shows that the initial mass estimate by \cite{wollman1977} was already quite close to the
results obtained using the stars.
There is also a clear trend for a mass increase as a function of time visible.
While the early measurements using stellar radial velocities and proper motions all depend on
modeling the volume mass density and the three dimensional distant dependency of the velocity
dispersion, the more recent measurements rely on the orbital analysis of the innermost stars
(predominantly S2).
Using the closest stars possible, the 
mass estimate may now be expected to settle around a value of 4.0 million solar masses with 
an uncertainty of about 0.2 million solar masses
\citep{Boehle2016, Parsa2017}.
As mass and distance are determined simultaneously from stellar orbits, future improvements of the 
uncertainties must involve further improvements of the positional and spectroscopic measurement accuracy.

\subsection{Size}
\label{Size}

Knowing the distance and the mass we can now discuss the observed 
and estimated sizes of the black hole at the Galactic Center. 
Knowing the size then allows us then to calculate the mass density and 
compare it to mass densities expected for a SMBH of the given mass.
Hence, we can verify how compact the massive object at the center is.
As a size we may refer to a measurable diameter of the smallest possible 
region that contains the mass of the black hole. For a black hole the
event horizon qualifies for such a designation
although it can be regarded a unobservable by definition
(see sections \ref{Introduction} and \ref{predictions}).
But one can get very close to it
(see sections \ref{Horizon} and \ref{conclusion}).
The core radius of hypothetical dense clusters (Plummer models 
\footnote{A Plummer 3-dimensional density profile with the Plummer radius $r_0$ is given by
$\rho=\rho_0 (1+\frac{r}{r_0})^{-\alpha}$. Functions of this form are often used as models to 
describe the density profile of dense stellar clusters.}
with an exponent of $\alpha$=5) 
constrained by the peribothron distance of the S2 orbit
that could hypothetically still explain the 
total mass concentration would have  a radius close to 0.22mpc.
Such a cluster would have a very short live time of less than $10^5$~yr. 
The radius of a neutrino ball composed of degenerate 17~keV neutrinos, for example, 
is difficult to reconcile; 
as summarized in section \ref{FermionBall} on alternative models to the black hole.

\begin{figure*}
\begin{center}
\includegraphics[width=12cm]{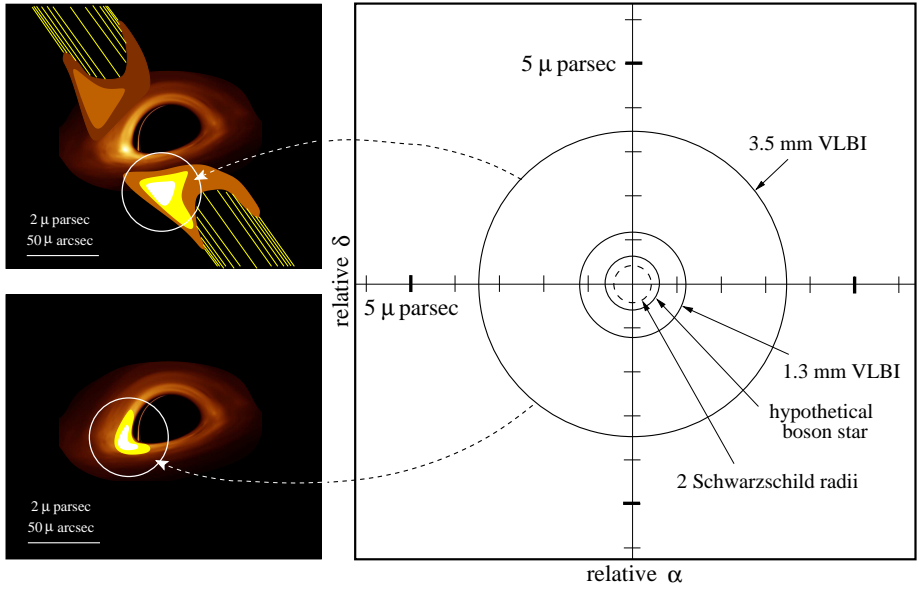}
\end{center}
\caption{A comparison of size scales at the Galactic Center near Sgr~A*. 
We show results from theoretical predictions.
On the right we depict different size estimates and scales. 
In this section of the image we assume that the resolution elements 
are centered on the position of the SMBH. However, on the left we 
are centered observationally on the peaks of the brightness distributions.
The white circle highlights the brightest structure in the image and indicates a 3.5mm (full width at half maximum) VLBI beam.
To the upper left we show the the expected image in the case of a jet foot-point case and to the lower left the accretion disk case.
The VLBI beams have been calculate assuming a 4500~km baseline.
For a rotating helical jet the jet brightness profiles are highly asymmetric and due to 
boosting and light bending effects for medium to low 
inclinations the receding counter-jet will dominate the jet section pointed towards the observer,
here the lower right site of the jet foot-point panel; see \cite{Dexter2012a}.
Approximate size scales for the SgrA* case are given in the lower left hand side of each panel.
\label{fig4}}
\end{figure*}

In Fig.~\ref{fig4}, we show a comparison of size scales at the Galactic Center near Sgr~A*. 
\cite{shen2005}
report on a radio image of Sgr A* at a wavelength of 3.5mm, demonstrating that its size is 1 AU (or 4.9 $\mu$pc).
When combined with the lower limit on its mass (\cite{Reid2003}, see also \cite{reid2003b}), 
the lower limit on the
mass density is 6.5$\times$10$^{21}$ \solm pc$^{-3}$, which provides the most stringent evidence 
to date that Sgr A* is a SMBH. 
The black hole mass density for 4 million solar masses in a sphere of the corresponding Schwarzschild radius
is 1.63$\times$10$^{25}$ \solm pc$^{-3}$.

{\bf 1.3mm VLBI source structure} smaller
than the expected apparent size of the SgrA* black hole event horizon has been observed by \cite{doeleman2008},
suggesting that the bulk of SgrA* emission may not be centered on the black hole, but arises in the
surrounding accretion flow. As shown in Fig.e,f this can be explained either as a compact jet foot-point
or as the Doppler enhanced, approaching side of an - at least temporary luminous - accretion
flow or disk. The intrinsic size of Sgr A* is equal to 37$^{+16}_{-10}$ $\mu$as ($\sim$1.5$\mu$pc; see Fig.~\ref{fig4}).
The corresponding 3$\sigma$ upper limit 
of the source size at 1.3 mm, combined with a lower limit to the mass of Sgr A* of 4$\times$10$^5$ \solm
\citep{Reid2003},
yields a lower limit for the mass density of 9.3$\times$10$^{22}$ \solm pc$^{-3}$. 
\cite{fish2011, doeleman2009}
\cite{doeleman2009} show that source components orbiting the SMBH can be detected even with non-imaging EHT data sets,
and \cite{fish2011, fish2016} confirm strong evidence for time-variable changes in SgrA* on scales of a few
Schwarzschild radii using EHT observations. 
 With these refined analysis tools and successful
pre-cursor observations, the prospects for directly probing the event horizon with the EHT
are excellent, and present exciting opportunities to constrain physics and dynamics at the
black hole boundary, as well as potential tests of GR, in conjunction with NIR observations \citep{Psaltis2016}.
Also in Fig. \ref{fig4} we show for comparison a typical size of a few Schwarzschild radii in the hypothetical case of a
boson ball as summarized in section \ref{FermionBall} on alternative models to the black hole.

\subsection{Spin, Inclination, and frame-dragging}
\label{Spin}

The spin\footnote{For black holes the spin can be characterized by via the angular
momentum parameter $a=J/Mc$, where
$J$ is the angular momentum and $M$ the mass of the black hole. It has the dimension of a length and
lies in the interval $a^2 \le (GM/c^2)^2$.
This results into the dimensionless parameter $a^*=ac^2/GM$, which now lies between 0 and 1.
Non-rotation black holes have $a^*=0$ and maximally rotation black holes have $a^*=1$.
The spin is usually determined from modeling spin dependent quantities, like light curves of 
orbiting hot spots or observing jets 
(in section \ref{jet}) or searching for a black hole shadow (in section \ref{shadow}).}
is one of the fundamental quantities that characterize the black hole. 
In combination with the inclination its value is a means to probe the observability
of predicted effects (see sections \ref{jet} and \ref{shadow}).
\cite{eckart2006}
report polarization measurements of the variable near-infrared emission of SgrA* 
and investigate the physical processes responsible for the variable emission from SgrA*.
The authors find that the variable NIR emission of SgrA* shows highly polarized flux density excursions
supporting that the  NIR emission is non-thermal and could come from a jet or temporary disk model.
In comparison to an orbiting spot model they find that the variability is consistent with a spin parameter of
a=0.5 and appreciably large inclinations
\citep[see also][]{Shcherbakov2012, vincelt2011, Broderick2006, trippe2007}.
Partially ordered variable magnetic fields are also supported by recent 1.3mm VLBI measurements with the 
EHT \citep{Johnson2015}.
\cite{meyer2006} model their NIR polarimetry data successfully with a combined spot/ring model. 
They conclude that the inclination $i$ of the spot orbit must be larger than $20^o$ 
and the dimensionless spin parameter of the black hole is derived to be a$>$0.5. 
\cite{Meyer2007} find that their observations support an approximately constant mean 
polarization angle of 60$^o$$\pm$20$^o$ between 2004-2006.

The spacetime frame-dragging (Lense-Thirring) effects near a rotating 
compact object can be of
conceptual and also considerable astrophysical importance for the 
models of black holes surrounded by orbiting material, such as an 
accretion ring or a disk-like structure \citep{barbour1995}.
Naturally, these effects are the strongest near a rotating black hole, 
where the ergospheric region develops. 
For compact objects like stars probing the gravitational field of the SMBH,
the orbital period $P$ will depend on the black hole mass $M$ and the long axis $a$ of the orbital ellipse
and the relativistic effect of the high mass concentration will result in an angular advance of the 
peribothron (periapse for black holes) angle $\omega$,
\begin{eqnarray}
P \propto a^{1.5}M^{-0.5}~~~~~~~~~~~~\\
\omega \propto  a^{-1}M(1-e^2)^{-1}~.
\end{eqnarray}
In addition, precession of the orbits will be affected by the Lense-Thirring effect and 
the quadrupole moment of the black hole. The characteristic timescales for these three effects that dominate 
are $t_M$ (orbital precession), $t_J$ (Lense-Thirring precession), and $t_Q$ (quadrupole 
precession). They depend on successively increasing powers of the orbital long half axis and 
eccentricity factor $(1-e^2)$ and a decreasing power of the black hole mass 
\citep{Merritt2010a,Merritt2010b,psaltis2015a,Psaltis2016}.
\begin{eqnarray}
t_M \propto  a^{2.5}M^{-1.5}(1-e^2)~~~~\\
t_J \propto a^{3.0}M^{-2.0}(1-e^2)^{1.5}\\
t_Q \propto a^{3.5}M^{-2.5}(1-e^2)^{3}~~
\end{eqnarray}

This shows that the objects of very eccentric orbits ($e$ approaching 1) close to the black hole mass are preferred
and the higher order quantities like the precession due to Lense-Thirring and quadrupole moment become 
successively smaller.
Specific precession frequencies induced by the frame dragging could 
for instance be detected in motion of stars.
The number of stars available to prove relativistic effects within the central
arcsecond (about 10$^5$ Schwarzschild radii) can be estimated to be several 
hundred for a NIR brightness larger than 21$^{st}$ magnitude 
in K-band at 2$\mu$ wavelength and about 40 stars 
between 18$^{th}$  and 16$^{th}$ magnitude in the K-band
\citep{Sabha2012,Do2009,Rubilar2001,Jaroszynski1999}.
Relativistic effects are strongest close to the supermassive object.
Here measurements potentially also allow us to distinguish between a SMBH and a boson star (see~\ref{BosonBall}).
If we restrict ourself to the central 1000~AU we can estimate the number
of test objects that are available to probe the relativistic effects.

We estimate the amount of stellar mass $M_*(a)$ surrounding SgrA* 
inclosed within a circular orbit of semi-major axis $a$ as
\begin{equation}
\frac{M_*(a)}{M_{SgrA*}} = \left(\frac{M_*}{M_{SgrA*}}\right)\left(\frac{a}{a_0}\right)^{3-\gamma}~~.
\end{equation}
Here $\gamma$ is the exponent of the stellar number density distribution around the center.
We assume that within $a_0=1~pc$ a total mass of $M_*=$10$^6$\solm ~~~is 
contained in stars with typically one solar mass \citep{Merritt2010a}.
The results are summarized in Tab.~\ref{tab:numbers}.
Here it becomes evident that the number of objects strongly varies with the
value of $\gamma$.
We have used a value of $\gamma$=2 \citep{Merritt2010a}, close to the value of $\gamma$=7/4 
for relaxed stellar clusters around SMBHs \citep{BahcallWolf1976}.
We also used values for flatter distributions with  $\gamma$$\sim$1.2  
and $\gamma \sim$1.0  found for early and later type stars within the 
central arcseconds around SgrA*
\citep{Buchholz2009,Schoedel2007}.
If in Tab.~\ref{tab:numbers} 
the number of objects within the central 1000~AU gets close to or even drops well below unit.
In addition, these numbers can be looked upon as upper limits, for scattering
events may help to empty the central region around the SMBH.
In Tab.~\ref{tab:numbers} we have also listed the expected number of stars observed 
in the NIR K-band within the magnitude interval of K=18-19 \citep[see e.g.][]{Sabha2012}.
These will be suitable for NIR interferometry with GRAVITY at the VLTI (see section \ref{futureobs}).

\begin{table}[htb]
\caption{Number of stellar objects within 1000~AU of SgrA*}
\begin{center}
\begin{tabular}{ccccc}\hline \hline
$\gamma$ & N$_{stars}$ &N$_{msP}$ &N$_{nP}$ \\ \hline 
2.0 & 5000  & 5       & 0.5 \\
1.2 & 67    & 0.67    & 0.067\\
1.0 & 24    & 0.24    & 0.024\\ \hline 
2.0 & 6     & -       &- \\
1.2 & 0.08  & -       &- \\
1.0 & 0.03  & -       &- \\
\hline \hline
\end{tabular}
\end{center}
\small
Approximate number of stars, 
millisecond pulsars $msP$,
and 
normal pulsars $nP$ 
with distances to SgrA* of less than 1000~AU.
This corresponds to a radius of  0.125'' or 4.7~mpc.
Using a value of $M_*=10^6$\solm for the central paresc  we
derive for different values of $\gamma$ the number of solar mass stars (second column in the top three rows) 
and stars with a 2$\mu$m wavelength brightness in the magnitude interval K=18-19 
(second column in the bottom three rows).
Using the estimate 
of 100 normal and 1000 millisecond 
pulsars within the central parsec \citep{Wharton2012,psaltis2015a,Psaltis2016}.
we derived the corresponding values for the centra 1000~AU in cloumns 3 and 4. 
\label{tab:numbers}
\end{table}

\begin{table}
\caption{The inner radii of the stellar cusp for different stellar  
populations $M$ calculated according to Eq.~\ref{eq_inner_radius}.}
\centering
  \begin{tabular}{cccrrr}
  \hline
  \hline
   Stellar $M$ & $C_{M}$ & $\gamma_{M}$ & {\bf  $r_{1,M}\,[\rm{pc}]$}& {\bf  $r_{1,M}\,[\rm{AU}]$}& {\bf  $r_{1,M}\,[\rm{mas}]$}\\
population $M$ &         &              &                            &                            &                             \\
  \hline
  MS                     & $1$      & $1.4$          &  $2\times  10^{-4}$ & 41  & 5       \\
  WD                     & $0.1$    & $1.4$          &  $7\times  10^{-4}$ & 144 & 18      \\
  NS                     & $0.01$   & $1.5$          & $20\times  10^{-4}$ & 412 & 50       \\
  BH                     & $0.001$  & $2$            &  $6\times  10^{-4}$ & 124 & 15       \\
  \hline
\end{tabular}
\label{tab_inner_radius}
\end{table}

While in Tab.~\ref{tab:numbers} the number of pulsars is assumed to be a constant fraction
(see table caption)
of stars distributed with the power-law index $\gamma$, more detailed calculations can be 
performed that allow different kinds of stars to have different power-law exponents.
Similarly \cite{HopmanAlexander2006} calculated the  inner radius where the stellar cusp 
ends in a statistical manner (low  probability of detection) for different stellar 
components (general  notation $M$): main-sequence stars (MS), white dwarfs (WD), 
neutron  stars (NS), and black holes (BH). The general relation may be  expressed as,

\begin{equation}
  r_{1,M}=(C_{M} N_{h})^{-1/(3-\gamma_{\rm{M}})} r_{\rm{h}}\,,
  \label{eq_inner_radius}
\end{equation}

where $N_{\rm{h}}$ is the total number of MS stars, $C_{M} N_{h}$ is  the total number 
of stars of type $M$ within the radius of influence  of the black hole 
$r_{\rm{h}}=GM_{\bullet}/\sigma^2$ ($M_{\bullet}$ is  the black hole mass and $\sigma$ 
is the stellar velocity dispersion),  and $\gamma_{M}$ is the power-law exponent 
for stellar type $M$.  

According to the analysis of \cite{HopmanAlexander2006} the total  
number of MS stars within the radius of the gravitational influence  
$r_{h}=2\,\rm{pc}$ is $N_{h}=3.4\times 10^6$. 
Table~\ref{tab_inner_radius} summarizes the inner radii of the cusp for different components. 
The range of power-law exponents \cite{HopmanAlexander2006}
obtained in their dynamical calculation for different stellar populations (see Tab.~\ref{tab_inner_radius})
are included in the range of exponents covered in Tab.~\ref{tab:numbers}).
The fact that for pulsars the number of objects gets close to unity or actually drops below
unity within 1000~AU is consistent with the inner cusp radius of about 400~AU for neutron stars. 
According to Tabs.~\ref{tab:numbers} and \ref{tab_inner_radius},
the number of  objects within the central 1000~AU (inside the peribothron of S2 star)  
gets close to or drops below unity. This indicates that it is very unlikely that 
in there a star can be found and used to probe relativistic effects.

Specific precession frequencies induced by the frame dragging could also be 
detected in the time variable signal from SgrA* at the Galactic Center 
(i.e. precession of temporary accretion disks
and the corresponding modulation in the flux density), 
but also from a number of (more distant to us) active galactic nuclei
\citep[see also][]{Wu2016}.
The observation of these signatures could provide us with independent 
evidence for the presence of a rotating black hole in these objects. 
Solutions of Einstein equations for self-gravitating disks or rings with 
or without a central black hole have been found in the past
\citep[see e.g.][]{bardeen1971, will1974, neugebauer1993}.
Self-gravitating disks around relativistic spheroidal configurations 
have also been considered - however, those would be massive. They are important
for extragalactic SMBHs, but they are probably not of 
relevance in the case of the Galactic Center. 
Both the angular momentum of the black hole 
and the angular momentum of the surrounding disk contribute to the total dragging. 
If the black hole rotates slowly and the disk has sufficient mass, 
the maximum of the dragging effect is located close to the center of the disk 
rather than at the horizon \citep{karas2004}.
A possible observational consequence is that the light rays near a 
self-gravitating disk are significantly distorted, which results 
in the change of the spectrum compared to the case when the disk 
self-gravity is ignored. 
Another consequence is that trajectories of massive bodies near the disk 
are attracted to it. Similar to the effects on massive particles, 
the frame-dragging effects can be revealed in the shape of magnetic 
lines of force that are also affected by rotation of the black hole, 
twisting the magnetic structures in its close vicinity \citep{karas2012}
In the case of the Galactic Center the observable  
frame-dragging effects near SgrA* are most likely limited to the stellar motion 
- if stars close enough to SgrA* can be found - and characteristic quasi-periodic 
modulations of the light curves.

\subsection{Orientation}
\label{Orientation}
The orientation angle $\theta$ of the inclined spin axis on the sky may be linked to the 
directions under which jet or wind interactions with the surrounding interstellar medium are claimed
\citep[see right side of Fig.~\ref{fig1}, section \ref{jet} and discussion by][]{shahzamanian2015, li2013,YusefZadeh2012a, eckart2006, eckart2006b,morris2004}.
The theoretical study of outflows from RIAF was performed by \cite{Yuan2012},
Based on NIR polarimetry data taken over a time range of 8 years \cite{2015Shahzamanian}
find typical polarization degrees that are on the order of 20\%-10\% and a preferred polarization angle of 13$^o$$\pm$15$^o$. 
The emission is most likely due to optically thin synchrotron radiation, and the preferred polarization angle is very 
likely coupled to the intrinsic orientation of the Sgr A* system, i.e. a disk or jet/wind scenario associated 
with the SMBH. 
\cite{2015Shahzamanian} conclude that if the polarization properties are linked to structural features, then the data imply 
a stable geometry and a stable accretion process for the Sgr A* system.
The polarization position angle taken as a measure of orientation 
of the SgrA* system on the sky
matches quiet well with the relative location of the mini-cavity 
(at an angle of about $\theta$=193$^o$, i.e., $\theta$=13$^o$+180$^o$)
that may be due to the interaction of a nuclear wind from Sgr A* with 
the mini-spiral material \citep[see also][]{Rozanska2017}.
The cometary tails of sources X3 and X7 reported by \cite{muzic2010}
lie within this range of angles and can be taken as a strong observational support 
for the presence of a fast wind from Sgr A*.

Modeling the polarized light curves with a relativistic disk or hot-spot system 
results in model dependent information on the orientation of such a system on the sky.
Using near-infrared polarimetric observations and modeling the terms of an orbiting spot,
\cite{Meyer2007}
constrain the three dimensional orientation of the Sgr A* system.
They find that the position angle of the equatorial plane normal is in the range 60$^o$-108$^o$ (east of north) 
in combination with the large inclination angle. 
This is in agreement with the orientation for emission from disk components with
magnetic field lines perpendicular to the disk as described by \cite{2010Zamaninasab}.
In \cite{2010Zamaninasab}, the authors also find significant evidence from 
emission of matter orbiting a SMBH.

A range of orientation angles of the SgrA* system on the sky may be linked to
jet or wind phenomena that are observed within the central stellar cluster.
A detailed summary and discussion of these phenomena is given by \cite{2015Shahzamanian}.
Without a clear kinematic evidence for a jet or wind and its connection to SgrA*,
no firm conclusions can be drawn on how the system is oriented in the sky and how the 
polarization and modeling information can be interpreted in this context.
\cite{psaltis2015b} and \cite{Psaltis2016} discuss observations with the Event Horizon Telescope at 1.3 mm 
which have revealed a size of the emitting region that is smaller than 
the size of the black-hole shadow. They argue that this can be reconciled with 
the spectral properties of the source, 
if the accretion flow is seen at a relatively high inclination (50$^o$-60$^o$). 
\cite{psaltis2015b} and and \cite{Psaltis2016}
claim that such an inclination makes the angular momentum of the flow, 
and perhaps of the black hole, nearly aligned with the angular momenta of the 
orbits of mass-losing stars that lie within about 3'' from the black hole and are the main source for 
the putative accretion stream onto SgrA* \citep[see also appendix by][]{muzic2007}.

\subsection{Spectrum}
\label{Spectrum}

\begin{figure*}
\begin{center}
\includegraphics[width=12cm]{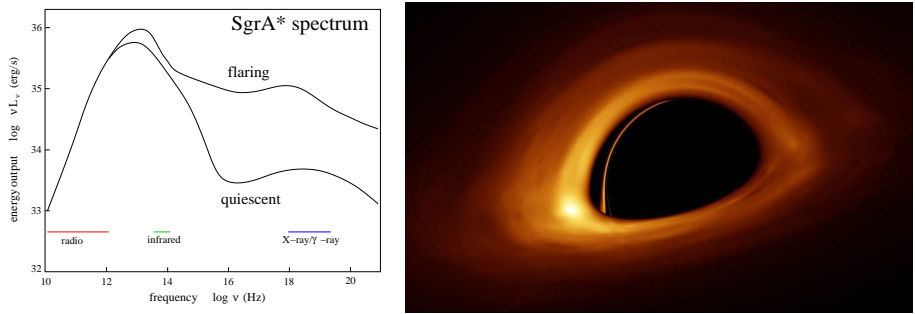}
\end{center}
\caption{
{\bf Left:}
Sketch of the broad band electromagnetic spectrum of SgrA*.
Data can be obtained in the accessible and useful radio, infrared and X-ray/$\gamma$-ray windows.
The black curves represent a model spectrum during quiescence and during a bright flare of SgrA*.
The model spectrum matches the observed data well and predicts what is expected in the
non-accessible regions of the spectrum.
{\bf Right:}
Model image of the shadow of the black hole
(see \cite{bursa2007} more details on the calculations).
Relativistic effects brighten the right left side and 
dim the right side of the temporary accretion disk surrounding the black hole at the center. For suitable 
inclinations of the system, relativistic effects (e.g. light bending and abberation) produce 
a dark region close to the position of the black hole and covering about 
a fourth of the shown image section. For a non-rotation black hole of 
$\sim$4.3$\pm$0.3$\times$10$^6$\solm, the image has a size of 
about 9.4$\times$15.4 Schwarzschild radii
corresponding to an angular size of about 160$\mu$as$\times$100$\mu$as. 
\label{fig5}}
\end{figure*}

SgrA* shows variable emission from the radio to the X-ray domain.
The overall spectrum of SgrA* may be used to derive general properties of the 
immediate physical surrounding of SgrA*.
As in the case of stellar accreting black holes
the spectral properties of this region for accreting SMBHs
are intimately linked to the accretion state and the black hole location in
a fundamental plane expressing that state
\citep[e.g.][]{Markoff2011, Markoff2005, VasudevanFabian2007}.
On the left side of Fig.\ref{fig5} we show a sketch of a spectrum of SgrA* in quiescent and in 
a flaring state from the radio to the X-ray/$\gamma$-ray domain.
\cite{2003baganoff} show that the X-ray emission at the position of Sgr A* is extended, with an intrinsic 
size of $\sim$1.4" (FWHM), consistent with the Bondi accretion radius for a several million solar mass black hole. 
However, SgrA* is much fainter than expected at all wavelengths, especially in X-rays.
\cite{2001baganoff} report the first discovery of rapid X-ray flaring from SgrA* using the Chandra 
observatory in the (0.5-7 keV)-band implying that this part of the X-ray emission is due to accretion of 
gas onto a SMBH.

\cite{2012eckart}
report on new simultaneous observations and modeling of the millimeter, near-infrared, and X-ray flare emission of the source SgrA*.
The authors study how and if the concept of the adiabatic synchrotron source expansion can be applied to the variable emission of SgrA*.
Source component sizes are typically around one Schwarzschild radius and the peak or the flare emitting spectrum 
lies around 300-400~GHz or just short of 1~THz. The bulk of the emission can be explained 
through Synchrotron-Self-Compton (SSC) scattering from these synchrotron sources into the X-ray domain \citep[see also][]{YusefZadeh2011,YusefZadeh2012b}.

Modeling of the light curves shows that the sub-mm follows the NIR emission with a delay of about 
three-quarters of an hour with an expansion velocity around 10$^{-2}$ of the speed of light.
In the radio flare the variability often shows signs of adiabatically expansion of these synchrotron source components.

Assuming that the bulk of the emission from SgrA* originates from accretion of matter 
provided by streamers and strong stellar winds in the
central stellar cluster, SgrA* provides an outstanding case for a radiatively inefficient 
accretion flow (RIAF), a favorite model for the accretion of matter onto SgrA*
with accretion efficiencies well below the standard thin-disk accretion flow
\citep{Quataert2003,YuanNarayan2014}.
SgrA* radiates at about 10$^{-9}$ times the Eddington luminosity which is the maximum that can be 
achieved for spherical accretion onto a massive black hole.
Hence, SgrA* is the weakest black hole accessible to detailed investigations
\citep{markoff2007}.
While nearly spherical accretion occurs if the angular momentum content of accreted 
material is negligible, in another limit one ignores radial transport of the material 
and considers an axially symmetric toroidal configuration that preserves permanent 
rotation about the symmetry axis. 
The treatment of self-gravitating discs was originally introduced in Ostriker's Newtonian 
equilibria of uniformly rotating, polytropic, slender rings \cite{ostriker1964}.
 Later, self-gravitating configurations with realistic equation of state and opacity 
were constructed and the basic formalism for self-gravitating black hole accretion discs 
was given in GR \citep{bardeen1973}.
Self-gravity has global consequences on the disk shape: it influences the location of 
its inner and outer edges, as well as the disk geometrical thickness and their vertical structure.

Radiatively inefficient accretion flows (RIAFs) are believed to power SMBHs 
in the underluminous cores of galaxies \citep{2014Moscibrodzka}.

Such black holes can typically be associated with compact radio sources with flat or 
inverted cm-/mm-radio spectrum as it is the case for SgrA*.
Using three-dimensional general relativistic magneto-hydrodynamics (MHD) accretion flow simulations,
\cite{2014Moscibrodzka}
can show that the radio to X-ray properties of SgrA* can very well be explained 
though a RIAF model under the conditions found at the center of the Milky Way.
In this case the X-ray emission is very sensitive to the electron heating mechanism in 
the immediate surroundings of the SMBH.

\subsection{Variability}
\label{Variability}

Given that the potential accretion  of stellar winds and mini-spiral gas may be a turbulent process,
one can expect that this will result in variable accretion and hence variable flux densities from the SMBH.
Alternatively the magnetic field strength and configuration may be affected by the interactions with the 
interstellar medium which would also result in strong variability.
\cite{2000falcke} present a summary of the wavelength dependent variability information for SgrA*
from the radio to the X-ray domain.
The authors found that synchrotron and synchrotron SSC provides an excellent fit 
to the entire broad band data.
The SSC process also efficiently explains the strong variability of SgrA* seen at X-ray bands.
If SgrA* has a jet, then it must be short or of very low surface brightness such that it 
naturally satisfies the 
resolution and sensitivity limits set by the interstellar scatter and VLBI experiments. 

\cite{2012witzel} characterize the statistical properties of the NIR variability of Sgr A*. 
They find that it is consistent with a single-state process, hence, forming a power-law 
distribution of the NIR flux density measurements. 
The authors also show that it is difficult with the current total power 
continuum data to decide on the existence of 
e.g. a quasi-periodic signal that is significant above the expected from a red-noise random process.
\cite{2015neilson} present a statistical analysis of the X-ray flux measurements of Sgr A* 
obtained with the Chandra observatory
describing SgrA* as a composite of a stationary source and a variable component
as expected for the inner section of an accretion flow. 
\cite{2015neilson} find that the variable component contributes about 10\% of the overall quiescent flux.
\cite{2014barriere} report the detection of SgrA* at energies of up to 79~keV using the NuSTAR observatory.
SgrA* continues to show strong variability at these energies as well and there is apparently not
sign for a cutoff in variability towards high energies.
Variable emission at high energies must result from inverse Compton scattering or from
highly efficient continuous particle acceleration and high magnetic field strengths 
as the synchrotron cooling time is of the order of 1~s.
\cite{2014barriere} highlight that with the variability time scales and the total energy 
emitted, the model-dependent location of the flares 
must be as close as  about 10 Schwarzschild radii from the black hole. 

\subsection{Does Sgr A* have a jet or an outflow?}
\label{jet}

Many (if not all) SMBHs in extragalactic low luminosity active galactic nuclei
are associated with jets that are most likely launched from their accretion disks.
Hence, the assumption is that SgrA*, although it is not an active galactic nucleus but is 
located at the lower end of the luminosity distribution
\citep[e.g.][]{eckart2012b, contini2011},
may have a jet or at least a strong wind, too.
Often the {"Fermi bubbles"} are quoted as potential observational signatures of strong outflow 
from hot accretion flow in the Galactic Center
\citep{Dobler2010, Su2010, Mou2014, Mou2015}.
These bubbles extend to about 50$^o$ above and below the Galactic plane, 
and exhibit a width of about 40$^o$ in Galactic longitude.
They are thought to be due to the interaction between the interstellar medium (ISM) and winds that have 
been launched from the hot accretion flow from SgrA* and the central starforming regions.

Proving the existence of a jet or strong wind would be very supportive 
for the existence of a SMBH.
Although observationally indicated  \citep{li2013,YusefZadeh2012a},
a jet or wind from SgrA* is expected but the evidences for it are not very clear
\citep[see discussion by][]{shahzamanian2015, eckart2006, eckart2006b,morris2004}.
The theoretical study of outflows from RIAF was performed by \cite{Yuan2012},
who show for the first time the existence of a strong outflow launched from the accretion flow.
(\cite{Blandford1999} only assume the existence of outflow.) 
\cite{Yuan2012} show that the inward decrease in the accretion rate which depends 
on the radius as $\frac{d}{dr}M_{acc} \propto r^s$ (where $s \sim 0.5-1.0$) 
is explained by the significant mass loss via wind.
MHD calculations predict a central plane and outflow region for the relativistic electron density distribution
\citep[e.g.][]{dexter2013, dexter2010, moscibrodzka2013}.
However, with decreasing radio wavelength the angular resolution of VLBI measurements is decreasing and
interstellar scattering is getting more severe as well \citep{markoff2007, britzen2015}.

Hence, this outflow region is difficult to be measured.
At mm-wavelengths the effects of interstellar scattering
can be overcome and the source intrinsic structure of SgrA* can be investigated.
However, steep spectra and low surface brightness makes it a challenge to search for a wind or jet 
from SgrA* at high frequencies.

\subsection{Event Horizon or surface?}
\label{Horizon}

In classical (non-quantum) GR, a black-hole horizon is defined as a 
null hyper-surface formed by light rays that are just on the verge between 
escaping to infinity and being trapped by the strong gravity. 
As a consequence, the horizon is of  rather ill determined nature in the sense 
that one has to know the entire history 
of light rays and the null structure of the spacetime to be able to determine the existence and the 
actual location of the horizon \citep[see][]{israel1987}).
Could SgrA* be supermassive but have no event horizon?
If such a possibility can be ruled out, this would speak 
in favor of it being a black hole.
Ruling out such a possibility would speak in favor of a black hole.
The presence of a jet implies a high accretion rate either onto an 
accretion disk, a hard surface, or an event horizen.
The presence of a jet origination from the vicinity of SgrA* would imply an at 
least temporally present accretion disk
that may maintain magnetic flux required for the jet confinement and launching near the black hole. 
Regardless of the jet launching mechanism, a minimum mass accretion rate would 
be necessary to power this scenario.
This has been discussed in detail for the M87 jet by \cite{broderick2015}.
In the case of SgrA*, such an accretion rate is provided by the surrounding mass-losing young stars.
Accreting this material onto a hard surface instead through an event horizon 
would result in considerable thermal near-infrared emission from the surface due to shocks
\citep[e.g.][]{narayan1998,verozub2006}.
The fact that SgrA* is sub-luminous even with respect to the Eddington luminosity, 
which results from the highest rate at which
matter can be spherically accreted onto black holes, implies the presence of a black-hole 
horizon rather than a hard surface.

It appears to be virtually impossible to unambiguously prove the presence 
of black-hole horizon by observing the electromagnetic signal from its presumed 
vicinity to such a horizon. Here, the absence of a radiation signal does not 
necessarily prove the absence of the surface of the body and the existence of the horizon
\citep[see the critical discussion of tentative proofs, like general
properties of accretion flows, dimness, absence of X-ray bursts, in][]{abramovicz2002}.
However, a very suggestive and almost convincing evidence of the strong gravity 
associated with stellar or supermassive black hole would be a detection of light encircling the black hole 
along a photon orbit and thus leading to multiple images that 
should occur with a precisely defined time delay. 
As can be seen by the communication between the Czech engineer Rudi Mandl and Albert Einstein 
\citep{renn2005, einstein1936} the concept of light bending in the context of 
high mass concentrations has been discussed at a very early stage.
\cite{bursa2007} conclude that if the delay in arrival time between the ``direct" and ``looped" photons can 
indeed be revealed in the light curves, this would provide a direct evidence for the existence of 
circular photon orbits. In this way, it will be possible to demonstrate the validity of an important 
prediction of GR in the regime of strong gravitational fields.

\subsection{Black Hole Shadow: proving GR and observing the photon sphere at the event horizon} 
\label{shadow} 

Very long baseline radio interferometry in the mm-wavelength domain promises to
image the immediate vicinity of the SgrA* SMBH in the foreseeable future.
It is expected that the immediate surroundings of the SgrA* SMBH is 
lit up by accretion processes.
Due to the bending of light by the black hole there is the possibility that a dark 
region might appear in this region.
\cite{Falcke2000} have first predicted the size of this so-called shadow for Sgr A* of about 30$\mu$as. 
(see right side of Fig.\ref{fig5}).
The exact size, location and observability of the shadow will depend on the 
inclination and spin of the black hole.
However, it is within reach of current or upcoming VLB interferometers in the mm-wavelength domain.
These measurements must take into account that the imaging at least at wavelengths less 
than 1.3mm is hampered by interstellar scattering which is also responsible for part of the 
flux and structural variability.
\cite{Gwinn2014} have detected substructure within the smooth scattering disk of the 
SgrA* mm-VLBI radio image as a consequence of refraction in the interstellar medium.
\cite{Rauch2015} have detected a secondary radio off-core feature associated with flux density 
variations at mm- and NIR-wavelengths.
\cite{Fish2014} and \cite{Lu2016}
present a procedure to mitigate the effects of interstellar scattering.
They show that a black hole shadow and a photon ring (if both are indeed present) can clearly be detected 
if one makes use of observations over multiple days and creates an image of the average quiescent emission.
\cite{broderick2014} and others claim that the EHT-observations provide the novel opportunity 
to test the applicability of the Kerr metric to astrophysical black holes. 
Simulations by several groups have been prepared for this black hole shadow 
\citep[e.g.][]{ricarte2015, broderick2014, Falcke2000}.
\cite{broderick2014} present the first simulated images of a radiatively inefficient 
accretion flow (RIAF) around SgrA* employing a quasi-Kerr metric that contains an 
independent quadrupole moment in addition to the mass and spin 
that fully characterize a black hole in GR. 
They show that these images differ significantly from the 
images of an RIAF around a Kerr black hole with the same spin and demonstrate the feasibility of testing the 
no-hair theorem by constraining the quadrupole   deviation from the Kerr metric with existing EHT data. 
However, they claim that at present, the limits on potential modifications of the Kerr metric remain weak.

\subsection{Pulsars at the Galactic Center}
\label{pulsars} 
A very efficient way to map out space-time in the vicinity of 
the central super-massive black hole is to find and track 
pulsars orbiting SgrA*. 
The central black hole spin and quadrupole moment 
could be measured with a very high precision
by the pulsars that are only close enough to the black hole.
Tracking the orbital motion of pulsars would allow us to test different theories of gravity
\citep{Psaltis2012}.
Although it is generally agreed on that there must be a large number of stellar remnants at the center
of the stellar cluster, i.e. in the immediate vicinity of the black bole,
it is not clear how large  the number of detectable pulsars in that region will actually be.
These number will depend on the star formation history in the overall region and on the efficiency 
of the dynamical processes that let remnants gather in the very center.
\cite{DexterOLeary2014}
explain the ``missing pulsar problem" by an intrinsic lack of ordinary 
pulsars. On the other hand (although the statistics is still poor), the discovery of a single magnetar 
within 0.1 pc may imply an intrinsic overproduction of magnetars 
close to the Galactic Center since the occurrence of magnetars 
in the standard Galactic stellar population is rare in comparison 
with normal pulsars. The overproduction of magnetars could take 
place due to either strongly magnetized progenitors or the 
top-heavy initial mass function (massive progenitors).
Magnetars are short-lived in comparison
with ordinary pulsars, which results in low likelihood of pulsar detections
( \cite{DexterOLeary2014} compare sensitivity limits of past surveys with luminosities of
known pulsars in their Fig.~2).

Estimates of the number of solar type stars, normal and millisecond pulsars are listed in Tab.~\ref{tab:numbers}.
These numbers imply that the likelihood of finding even a single pulsar in a region in which
the gravitational field of the central SMBH can be 
investigated is extremely small.
However, the magnetar PSR~J1745-2900 resulted in the discovery of radio pulses using the 
Effelsberg telescope. This discovery demonstrates that these objects are present in the 
central stellar cluster and that highlights the great value and the efforts 
that are being undertaken to find pulsars close to SgrA*
\citep{Spitler2014, Mori2013, Eatough2013}.
Alternatively, instead of looking for radio pulses it was proposed  that several neutron stars 
could be potentially detected via bow shocks due to the interaction of supersonic magnetized 
neutron stars  with ionized gas in the central parsec \citep{Zajacek2015, GianniosLorimer2016}.
\cite{Psaltis2016}
show that the results obtained from stars and pulsars can be ideally combined 
with those of shape and size of the shadow of the black hole promising a high accuracy 
test for the gravitational no-hair theorem.

\subsection{Any chance for charge?}
\label{charge}

As stationary black holes are characterized by the three quantities mass, angular momentum and charge, one may raise the question 
if there can be a significant electrical charge associated with the SMBH at the center of the Milky Way.
Although GR allows for a black hole to acquire electric charge, 
it is thought that astrophysical black holes are electrically neutral to very high precision. 
This is because of processes of selective charge accretion from the surrounding plasma. 
However, it is important to note that the surrounding material may acquire some non-zero 
net charge density, \cite{slany2013}, and so the interplay of electrically charged 
particles and magnetic fields in the strong gravitational field of black holes appears to be relevant. 
For example, irradiation of dust particles leads to a positive net electrical charge by photoionization. 
On the other hand, plasma electron and ion currents are continuously entering the 
grain surface, so the sign and magnitude of the equilibrium charge depend on the total 
currents that are absorbed and emitted from the grain surface. 
This is a complex process that depends on many parameters, and the resulting message 
is that electrostatic charge is among the essential parameters that control the dynamics 
of dust grains embedded in the surrounding plasma.
However, in the Galactic Center this effect may be diminished greatly.
The existence of smaller dust grains at the center 
is probably limited due to intense UV and X-ray radiation field 
and they plausibly completely vanish on the scale of tens of gravitational radii in the hot accretion flow. 
On the other hand, the continuous formation of dust in stellar winds, in particular of AGB stars
\citep{YusefZadeh2017}, as well as the speculative 
asteroid or planet infall towards Sgr A* \citep{Zubovas2012} 
could in principle replenish the dust content even very close to the black hole.

Black holes without angular momentum are described by the Reissner-Nordstr\"om metric,
those with angular momentum and electric charge are described by a Kerr-Newman metric.
As the electrical force field is about 40 magnitudes larger than the gravitational field,
it is thought to be unlikely that significant netto charges can be accumulated with black holes.
There is the possibility that the black hole shadow disappears if an appreciable charge is accumulated
\citep{Zakharov2014}.
However, the shadow of a black hole is not a very clean observable, and inclinations and 
accretion phenomena may cause the disappearance (or washiness) of a shadow as well.
In 2013, Tsupko \& Bisnovatyi-Kogan 
gave a first detailed analysis of how the images from the surroundings of black holes change 
in the presence of plasma \citep{Tsupko2013}.
This may imply that suitable observing frequencies may lie even above 230~GHz
\citep{Falcke2000} to avoid that the images are severly washed out.

\subsection{Gravitational wave signals from inspiraling objects}
\label{ringing} 
The merging event between SMBHs among each other and with other compact objects like 
will produce specific 
ringing signals in emitted gravitational waves \citep{Loeckmann2008, Aasi2013}.
In a strict sense the word ``ringing" is used to describe the ring-down
of the new SMBH formed in the aftermath of a SMBH binary merger.
For simplicity we use it here for the gravitational wave signal we receive
from inspiraling masses in general.
Comparison of these signals to theoretically 
predicted signals is a direct indication for the presence of a SMBH.
These signals may be the only radiation coming 
from the immediate vicinity of the black hole
(see section~\ref{subBHinGR}).
Extreme Mass Ratio Inspirals \citep[EMRIs; see][for a review]{2007CQGra..24R.113A}
have also been identified as effective probes for the Kerr metric of GR 
by the eLISA group \citep{Danzmann2015}.
However, the merging rates are very small and a high detection rate can only be achieved
by including a large local cosmological volume.
The recent measurement of a theoretically expected chirped gravitational wave signal  
using the Laser Interferometer Gravitational-Wave Observatory (LIGO) 
demonstrated the existence of binary stellar-mass black hole systems 
and the possibility to directly detect gravitational waves from a binary black hole merger
\citep{Abbott2016}. 
It was reported that at a distance of about 410~Mpc
in the system GW~150914 two very compact objects of about 36~\solm ~and 29~\solm ~merged.
These objects were probably black holes; however, exotic alternatives still need to be 
ruled out (see sections \ref{Alternatives} and \ref{subsection:antirealism}, 
\cite{Cardoso2016}, as well as a comment in \cite{LIGO2016}).
For most of the alternative theories, predictions for the specific shape of the ringing 
signal still need to be performed,
although at a higher frequency, these measurements validate the method proposed to search for 
supermassive binary black holes and support the possibility of detection gravitational wave ringing 
from such systems.
If there are IMBHs in the central stellar cluster 
\citep[e.g. the case of IRS13E][]{maillard2004, schoedel2005, fritz2010},
their interaction with SgrA* may produce a sufficiently strong signal if a merger event occurs.
However, the likelihood for this is rather low \citep{Aasi2014}.

Expected gravitational wave events from the Galactic center are supposed to originate in the 
inspiral of bodies with extreme ratio of masses -- supermassive black hole and a stellar remnant 
of a few solar masses (EMRI; see above).
Extreme mass ratio means that a merging body effectively acts as a test particle in the spacetime of the 
supermassive black hole. Before the final plunge, it is possible to infer the properties of the spacetime 
from the observed waveform, since it reflects the peribothron as well as 
Lense-Thirring precession for a Kerr black hole.
White dwarfs, neutron stars and stellar black holes are not subject to 
tidal disruption. Merging with a SMBH can hence result in a detectable amount of
gravitational wave luminosity (some 10$^{40}$~erg~s$^{-1}$)
over several days 
(see equation~\ref{eq_gw_luminosity} in Appendix \ref{App:ringing}).

\begin{figure}
  \includegraphics[width=12cm]{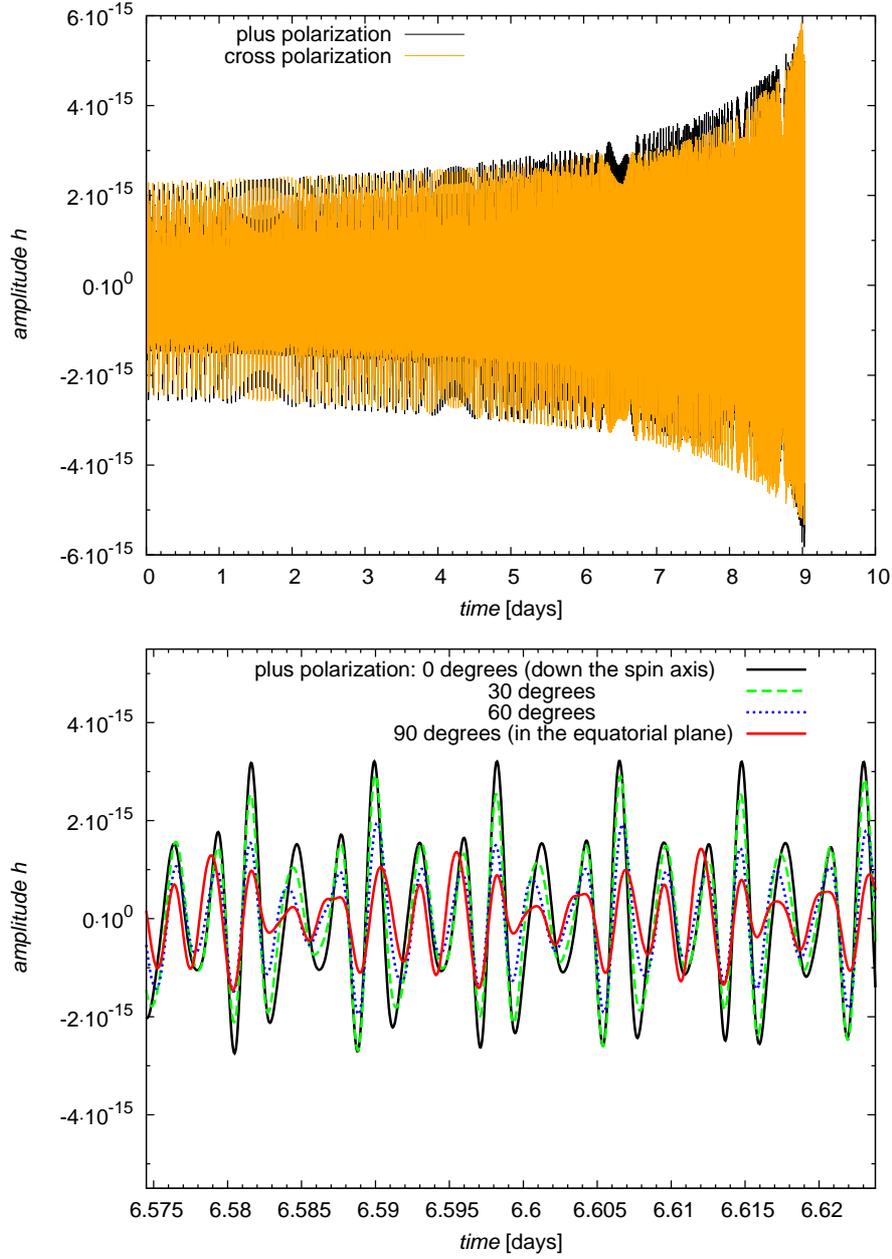}
  \caption{Top: A waveform for an EMRI event starting at t=0 days at four gravitational radii for 
$m_{\star}/M_{\bullet}=10^{-4}$ (black hole spin 0.998) as viewed at $30^o$  from the 
spin axis of the black hole. The event corresponds to a non-equatorial orbit inclined at $40^o$. 
Two polarization modes (plus and cross) are depicted by different colors. 
The data for the plot were taken from the simulations of \cite{Hughes2000,Hughes2001}.}
Bottom: Zoom into the waveform around t=6.6 days.
Different lines  represent a different viewing angle (see the key).
  \label{fig_emri}
\end{figure}

 The detection of an EMRI event associated with the Galactic center black hole would 
enable to precisely measure the mass and the spin of the black hole with an 
independent measurement. 
The waveform could also reveal deviations from the Kerr metric as well as a possible 
different character of a central compact object (e.g. a boson star). 
An exemplary waveform for an EMRI event for the ratio $m_{\star}/M_{\bullet}=10^{-4}$ 
and the black hole spin $J=0.998$ is depicted in Fig.~\ref{fig_emri}. 
The inspiral starts at four gravitational radii.

The EMRI events from the Galactic center will be within the detection sensitivity 
limits of the planned LISA and eLISA space-base interferometers that will be 
capable to detect the gravitational wave events with low frequencies in 
the range $\nu_{\rm{GW}}=0.1\,\rm{mHz}$--$1\,\rm{mHz}$. 
Although the likelihood to detect an EMRI event for the Galactic center 
is rather low during the mission lifetime, \cite{2004CQGra..21S1595G} estimate 
that LISA will be able to detect $\sim 2$ EMRI events of $1.4\,M_{\odot}$ compact 
objects (white dwarfs and neutron stars) per cubic Gpc per year for 
black holes with similar masses as Sgr~A*. 

A very rare but rewarding event would be the detection of an inspiraling pulsar emitting gravitational waves. 
Such an event would allow us to do two independent measurements: An electromagnetic based timing measurement,
comparing the results with the general relativity predictions, and the other,  a GW-based measurement - also in comparison with theory. 
With a mass ratio of $m_{\star}/M_{\bullet} \sim 2 \times 10^{-7}$ this experiment is at the sensitivity limit and 
would only be possible for the Galactic centre, since for  other nuclei, pulsars would 
be too weak to be detected in the radio.  Basically the likelihood of this event is mainly given by the 
likelihood of detecting a pulsar on a relativistic orbit. In principle such a pulsar would already be
emitting GWs with the flux depending on its semi-major axis (see equations in  the appendix~\ref{App:ringing}).

\subsection{Manipulative success}
\label{manipulative} 
We may ask if the concept of the ``manipulative success" presented 
in the framework of {\it Entity Realism} in section~\ref{PhilConcepts}
can be applied to SMBHs.
Of course, we cannot ``use" and ``manipulate" them, 
but this is the case for almost all phenomena studied in astrophysics.
The problem, however, is solvable statistically.
In this sense, we can study candidates for SMBHs in different environments 
(i.e. in nuclei of galaxies of different types and masses); hence, we can
quasi exploit them in a statistical way by observing samples of
SMBHs and SgrA* at the center of the Milky Way is one of them.
Examples for such studies are the 
scaling relations for SMBHs \citep{Graham2015, Ferremateu2015},
merger scenarii for gravitational waves detection \citep{Schnittman2013},
reverberation analyses of black hole masses \citep{Shen2015, Matsuoka2015},
recoil scenarii in galactic nuclei \citep{Markakis2015, Guedes2011}.

All of these phenomena need large compact masses as they are provided by  SMBHs.
It is, however, currently difficult to distinguish 
between an action induced by a SMBH or a very compact boson star (see section~\ref{BosonBall}).
To use the 
principle of ``manipulative success" we need observables that directly address the presence of an
event horizon. Here, jet launching mechanisms and QPO's (quasi-periodic oscillations) originating 
close to ISCOs (innermost stable orbits of matter orbiting black holes) appear to be suitable 
candidates especially for large SMBHs in the sky located in radio loud AGN. 
There, the advanced properties of mm-VLBI as expected to emerge from the EHT can be useful.

\section{Alternatives to the Black Hole Scenario} 
\label{Alternatives} 
If an alternative scenario described the observed phenomenon better, the consequences 
for astrophysics would be dramatic.
Depending on the model, this could effect the evolution of supernovae, 
the merging of binary black holes, evaporation of black holes, gravitational wave emission and gravitational 
wave background. The question whether the Galactic Center black hole is a black hole has far reaching 
astrophysical consequences because it is the best case for a black hole.
As discussed in section \ref{subBHandQT},
already for bona fide black holes quantum processes are of relevance.
The alternative scenarii that are discussed in the following fully rely
on quantum phenomena.
The so called ``fermion ball" and the ``boson star" scenarii,
discussed in the following subsections cover important ``dark particle matter" models that have 
been under discussion as an alternative to the central Galactic black hole.
A critical discussion on constraints on alternative models to SMBHs is given by \cite{miller2006}.
Here, we will briefly summarize their properties and discuss whether they are
suitable descriptions of the extreme mass concentration found at the position of SgrA*
(see also Fig.\ref{fig3}).
In sections \ref{gravastar} and \ref{macroquantumness}
we also mention the possibility of objects such as grava-stars and macro-quantumness as 
interesting concepts for interpretation and understanding the upcoming observational data.
Here, we will not refer to wormholes that have been shown to be unstable
by Homer Ellis in 1973 \citep{Ellis1973}.

\subsection{Fermion Ball}
\label{FermionBall}
The fermion ball as an attempt to explain large compact nuclear masses
observed at the centers of galaxies was introduced by
\cite{viollier1992} and \cite{depaolis2001}.
A motivation for the development of the neutrino ball scenario was
a decreasing radiative efficiency towards the center of the Milky Way. 
This could be linked to a resolved mass -- and therefore a decreasing 
gravitational potential near the very center.
Such a scenario would help to explain the low luminosity of Sgr~A*. 
In the specific case of a fermion ball, these objects are stabilized 
by the degeneracy pressure of e.g. neutrinos as the corresponding fermion candidates. 
Due to the Pauli principle, the degeneracy pressure of the fermions can be balanced 
by the self-gravity of a ball of degenerate fermions.
In this case, the non-relativistic Lane-Emden equation
can describe the relation between the mass $M$ and the radius $R$ of a
fermion ball, composed of fermions with mass $m$ and degeneracy $g$.
The maximal mass for a degenerate fermion ball, can be 
calculated in a general relativistic framework.
The maximum mass of such a degenerate fermion ball is given by the
Oppenheimer-Volkoff limit $M_{OV}$.
For a given fermion mass $m$, all objects heavier than $M_{OV}$
must then be black holes \citep{viollier1992}.

It is very difficult to explain all SMBHs like SgrA* at the low end and 
e.g. M87 at the upper end with the same 
fermion mass and degeneracy \citep{viollier1992}.
M87 \citep[e.g.][]{walsh2013} is one of the most 
massive central dark objects currently known,
with a mass of $\sim6.3\times10^{9}$\solm.
This implies a fermion mass around 17~keV/c$^2$. 
For a few objects one even infers masses that are almost a magnitude 
higher than that of M87
\citep{walker2014,ghisellini2009,ghisellini2010})
implying even smaller fermion masses.
A small fermion mass of 17~keV/c$^2$
will result in a fermion ball radius of about 10 light days, or
at the distance of SgrA*, about 8.3 mpc.
In this case, large sections of the S2 stellar orbit would be located inside the fermion ball.
The orbit of S2 would be expected not to be 
closed and would be subject to a significant Newtonian periabothron shift, 
since the extended fermion ball mass would be resolved by the orbit. 
This is due to the fact that 
the effective gravitational attraction the star is exposed to, 
changes towards smaller distances to the center.
Hence, it appears that one can exclude the possibility that all compact dark
objects at the centers of galaxies can be explained by a universal neutrino ball model.

On the other hand, the stellar orbit of S2 
\citep{gillessen2009a,horrobin2004,eisenhauer2003} and the fact that 
no significant Newtonian peribothron shift has been found
constraints the fermion mass and implies
that it should be higher than 400~keV/c$^2$
(e.g. \cite{viollier1992,Bilic2003}).
Hence, it is very difficult to explain all SMBHs like SgrA* at the 
low end and e.g. M87 at the high end with the same 
fermion mass and degeneracy \citep{viollier1992}.
This makes the concept of a neutrino ball not to appear very attractive.

In addition, in the case of SgrA* a neutrino ball could not account 
for the compactness of the SMBH observed at radio/mm wavelengths. 
Observations of X-ray and NIR flares from Sgr~A*
\citep{mossoux2015,neilsen2015,2014barriere,eckart2012, porquet2008, 
eckart2006b,ghez2004,eckart2003,genzel2003,2003baganoff,2001baganoff}
suggest that the emission comes from structures smaller than about ten
Schwarzschild radii of a $\sim 4\times10^{6}$~M$_{\odot}$ million solar
mass black hole. This is more than two orders of magnitude more
compact than the radius of a neutrino ball with a neutrino mass 
discussed in this context.

A further significant drawback of that scenario is that it 
does not explain what happens to the permanently in-falling baryonic matter.  
This matter will of course be 
trapped and concentrate at the bottom of the potential well.
This will unavoidably result into a seed black hole at some point. 
This scenario then defeats the purpose of having a ball of degenerate matter
(especially neutrinos; \cite{melia2001}).

\subsection{Boson Ball}
\label{BosonBall}

The boson star scenario is another model that could explain a very large and compact mass,
yet it is much more speculative than black holes.
It is a dark matter particle explanation that cannot easily be ruled out by the
present data. Such a ball of Bosons could be present in a very compact 
configuration that is difficult to distinguish from a black hole
in terms of compactness (i.e. size).
However, it is hard to understand how the
bosons managed to cool sufficiently in order to settle down into such a small
volume and do not form a black hole during that process \cite{maoz1998}.
Boson stars \citep{kaup1968} are supported by the Heisenberg uncertainty principle.  
\cite{ruffini1969} showed that -- e.g. for a boson mass of 1~GeV/c$^2$ -- a stable 
object of total mass of 10$^{-19}$\solm ~and 1~fm diameter could be formed. 
Obviously, the mass depends on the repulsive forces between the bosons.
If one wants to form objects with total masses
as large as they are found in galactic nuclei \citep{colpi1986},
then one must {\it ad hoc} introduce a hypothetical weak
repulsive force between bosons \citep{colpi1986}.
For a large range of hypothetical boson masses, they can have sizes of only several
times their Schwarzschild radii.  This is the prime reason why it is so difficult to clearly
distinguish observationally between compact boson stars and black holes as candidates
for supermassive objects as they are found at the nuclei of galaxies 
\citep[see also][]{torres2000, mielke2002, mielke2000}.

However, during its lifetime, even if a boson star had formed at the 
center, it should eventually
have collapsed to a black hole through accretion of the
abundant gas and dust in the Galactic Center.  Therefore we conclude that
similar to the fermion ball solution, a supermassive boson star is not an
astrophysically attractive explanation for the high mass concentration at the
center of the Milky Way. As for possibilities of definitely ruling out the
boson star scenario, simultaneous multi-wavelength
measurements of the emission from Sgr~A* 
\citep[e.g.][]{eckart2012}
will allow us to constrain the emission mechanism and therefore the compactness of the
emitting region around Sgr~A* even further.  Probably within the next decade
it will be possible to image the ``shadow" cast by the putative black hole
through deflection of light rays using global radio interferometry at
sub-millimeter wavelengths. Such an experiment will involve very long baseline
interferometry in the sub-mm regime \citep{melia2001,2000falcke}.
However, \cite{Vincent2015}
show that relativistic rotating boson stars can result in images that look
very similar to those expected from Kerr black holes, also revealing 
shadow-like and photon-ring-like structures. 
This demonstrates that it is very challenging to unambiguously discriminate the
presence of a black hole with an event horizon from other highly 
concentrated mass agglomerations.

The future for Galactic Center research lies in high angular resolution 
observations at all accessible wavelengths.
For radio wavelengths, progress will be made with VLBI at mm-wavelengths.
In the infrared wavelength domain interferometry is now possible with
large aperture interferometers that will - in the near future - allow us to
observe SgrA* with a resolution of a few milliarcseconds. 
These are the Very Large Telescope Interferometer, 
the W.M.Keck Interferometer, and the Large Binocular Telescope.
\citep{pott2005, eckartpott2006, pott2008} have carried out first
mid-infrared interferometric observations of a number of bright
10~$\mu$m sources within the central stellar cluster using MIDI
\citep{leinert1998} and the 47~m UT2/UT3 baseline. 
In these observations, fringes on the first Galactic Center source were
obtained on the stellar source IRS3.
In the very near future further infrared interferometer measurements 
in the Galactic Center area will be possible 
\citep{eisenhauer2008, eckart2010, eckart2012c, vincent2011}.

The alternative explanation of the central mass as a massive boson star
(\cite{torres2000, mielke2000, lu2003} and references therein)
is {\it severely challenged} by the 
good agreement between the measured polarized flare structure 
and the theoretical predictions 
\citep{Gillessen2006, eisenhauer2005, Broderick2005, ghez2004}
as well as the indication of a quasi-periodicity in the data.
If an {\it ad hoc} weak repulsive force between a hypothetical brand
of bosons is introduced, it appears to be possible to form massive,
compact objects with sizes of a few R$_S$ that
are supposedly supported by the Heisenberg uncertainty principle. 
However, it is a delicate process to form a boson star and 
preventing it from collapsing to a 
MBH despite of further accretion of matter, 
a non spherically symmetric arrangement of
forces as in the case of a jet or matter 
being in orbit around the center but well within the boson star.
Such a massive boson star scenario could already be excluded for the 
nucleus of MCG-6-30-15 \citep{lu2003}.
Here, of the K$_{\alpha}$ line emission of the
relativistically moving  plasma of the accretion disk sourcounding the object 
could be used to put constraints on the mass concentration.

The results from \cite{Vincent2015}
show that for relativistic rotating boson stars 
it may still be very difficult to discriminate a boson star from a SMBH.
In the case of a stationary boson star, the orbital velocity 
close to the $\sim$3~R$_S$ radius of the last stable orbit
is already $\sim$3 times lower than that of a 
Schwarzschild MBH \citep{lu2003}
and relativistic effects are severely diminished 
and further reduced at even smaller radii.
If the indicated quasi-periodicity is due to orbital motion 
then a stationary boson star can be excluded
as an alternative solution for SgrA*, since in this case 
one expects the orbital periods to be larger.
However, if pulsars on highly eccentric orbits close to SgrA* can be found, 
then the black hole versus boson star riddle may be solvable by measuring the
central massive object's quadrupole moment  \citep[e.g.][]{Psaltis2012}.
How Kerr black holes and boson stars may be different with respect to their 
quadrupole moments is described in detail in \citep[e.g.][]{herdeiro2014}.
Whether such a measurement is successful will also depend on the environment and orbital disturbances.

\subsection{Grava-stars}
\label{gravastar}

The grava-star model postulates a strongly correlated thin shell of
anisotropic matter surrounding a region of anti-de Sitter space. 
It has been proposed as an alternative to black holes first by \cite{mazur2001}.
\cite{broderick2007} discuss the grava-star model for SgrA* but conclude that 
present day astronomical observations rule out modifications of GR 
of the kind described by \cite{chapline2003} on all scales larger than the Planck length.

One observational consequence of the analysis presented by \cite{broderick2007} 
is that (in particular) in the case of SgrA*, a grava-star will not have had 
sufficient opportunity to cool if the accretion is continuous.
However, if mass accretion is predominantly done through 
stellar capture events during a rather transient rapid accretion, this would allow us  
SgrA* to cool over timescales between captures of the order of $10^4$ years.
While occasional stellar captures may occur, the current observational evidence
suggests a rather continuous mass accretion at least over the past $10^5-10^6$ years
given the large number massive Helium stars that lose $\sim 10^{-3}$\solm~yr$^{-1}$ 
into the deep gravitational potential well of SgrA*. 
The radiatively inefficient accretion flow associated with SgrA* 
(see section \ref{Spectrum}) suggests that only 
between $10^{-8}$ and a few times $10^{-10}$\solm~yr$^{-1}$ reach the SMBH
\citep{Broderick2006, Yuan2003, Narayan1995}.
However, a clearer distinction between the black hole and the grava-star case is probably within
reach using high-resolution VLBI observations in the future \citep{Sakai2014}.

\subsection{Macro-quantumness}
\label{macroquantumness}

There could also be yet unknown observational effects of a strength depending on 
whether the composition of the black hole is rather baryonic or not.
\cite{DvaliGomez2011}
propose that Galactic black holes could be quantum objects i.e.,
Bose-Einstein condensates of $N$ soft gravitons at the quantum critical point, 
where $N$ Bogoliubov modes become gap-less
\citep[see also][]{DvaliGomez2013, DvaliGomezLuest2013}. 
Thus, predictions from semi-classical physics, 
usually applied to describe black holes, might not provide the proper descriptions. 
The metric itself might become an approximate entity. 
In the following we assume that the black holes 
carry a quantum memory about their baryonic content.  
This is mainly based on the quantum picture and model independent arguments.  
In case of baryonic content,  this is usually referred to as ``baryonic hair". 
It is still under discussion what the observable macroscopic effects are  
\citep[e.g.][]{Cunha2015},
but one can already argue about the relative strength of these effects
\citep[see also][]{Dvali2016}:
It seems that observable macroscopic effects of the ``baryonic hair" 
are suppressed by the ratio  $N_B/N$, where
$N_B$ is the baryon number of a black hole, and $N$ is the occupation number
of gravitons, which is equal to black hole entropy, i.e.,  $N = M^2/ M_P^2$,
where $M$ is black hole mass and $M_P$ is the Planck mass.
One may assume that this ratio must be small for large
black holes even if most of their initial mass is baryonic.  In this case,
$N_B = M/m_b$ ($m_b$ being baryon mass) will be a very small number as
compared to $N$, if the black hole radius is much larger than $1/m_b$.
On the other hand, the effects of ``baryon hair" may be strong as baryons
interact stronger than gravitons. 
This leaves us with the conclusion that the observable effects from
baryonic hair are expected to be very strong for very small black holes 
(of mass $10^{14}g$ or lighter).
Therefore, it cannot be excluded with certainty that the conventional
classical no-hair description is valid to a good accuracy for black holes
that are much heavier than $10^{14}g$, provided they are formed 
as a result of collapse of predominantly baryonic matter. 
In this case, the effects from ’baryon hair’ will most likely be very 
small for massive black holes.
However, if one allows that  the heavy black hole content could be non-baryonic
and that a black hole can form from some exotic light particles 
(e.g. dark matter of light sort), the situation may be very different.
In case these light particles have a mass 
comparable to the inverse gravitational radius, one can show that
the observational effects are very strong for large black hole masses.
It is unclear what these super-light hypothetical particles could be.
It should, however, be pointed out that this assumption is not any more 
exotic than the assumption of boson stars for which the hypothetical 
existence is based on the postulated  existence of new particle species 
and interactions amongst them.

\section{Future progress in observations} 
\label{futureobs} 

%ttttt
The combination of existing and planned millimeter/submillimeter 
facilities into the Event Horizon Telescope (EHT)\footnote{http://www.eventhorizontelescope.org/}
will give a high-sensitivity, high angular resolution of better than 60$\mu$as.
In the more distant future the EHT array may be supported by a mm-satellite facility 
in space to increase the angular resolution by factors of a few 
(see Millimetron Space Observatory; a 10~m aperture cooled telescope \cite{Kardashev2007,Kardashev2014}).
In particular the phased Atacama Large Millimeter Array (ALMA)\footnote{https://www.eso.org/sci/facilities/alma.html},
in Chile
will substantially contribute to this development.
Over the current decade, this instrument will finally allow us 
to directly probe the event horizon of the black hole candidates SgrA* and M87.
This requires the deployment of submillimeter dual-polarization receivers 
and highly stable frequency standards.
Such an equipment will finally enable us to carryout VLBI experiments at 230-450 GHz
in oder to perform a detailed and direct search for the edge of black hole. 
At longer radio wavelengths, in the future the Square Kilometre Array 
(SKA\footnote{The Square Kilometre Array (SKA) will be the world’s largest radio telescope, 
with a collecting area of more than one square kilometre (one million square metres);
https://www.skatelescope.org/}) 
will be sensitive enough to detect,
pulsars in a close orbit ($P_{orb} < 1~yr$) around Sgr A* and most importantly
measure its timing properties, which would enable to test ``cosmic censorship conjecture" 
as well as the ``no-hair" theorem \citep{psaltis2015a, Psaltis2016,2015Eatough}.

At near-infrared wavelengths, the 
GRAVITY\footnote{https://www.eso.org/sci/facilities/develop/instruments/gravity.html}
experiment at the Very Large Telescope Interferometer (VLTI) 
of the European Southern Observatory (ESO) will allow us to carry out precision narrow-angle 
astrometry in the ten microarcseconds and interferometric imaging in the milliarcsecond regime. 
GRAVITY will allow us to use dual beam phase reference observations of SgrA*
and will therefore as a second generation VLTI instrument
enhance the sensitivity and accuracy far beyond todays limits.
GRAVITY will be able to measure the motion of the photo-center of the SgrA* image during flares. 
If the flares are linked to orbital motion of a hot-spot within a 
temporarily bright accretion disk, GRAVITY will see the periodic oscillations of the
centroid position.
Similarly, bright outflow components that separate from SgrA* can be detected and 
distinguished from components within the accretion disk.

Across the wavelength domain new instrumentation like the 
James Webb Space Telescope (JWST)\footnote{http://www.jwst.nasa.gov/}
or the Constellation X\footnote{http://constellation.gsfc.nasa.gov/}
and Xeus X-ray\footnote{http://www.rssd.esa.int/index.php?project=XEUS} 
satellite missions will provide sensitive 
information on the spectral shape and variability of SgrA*.
For instance SgrA* has not yet been detected in the mid-infrared wavelength domain 
longward of about 5$\mu$m wavelength. The high point source sensitivity and 
foreseeably stable point spread function of the JWST will allow for progress.
The combination of high sensitivity and imaging quality will also help to investigate the 
nature of faint X-ray flares that are currently difficult to measure due to low 
count rates and contamination of a prominent Bremsstrahlung source surrounding SgrA*. 
The investigation of faint flares and the comparison with flares at different wavelengths 
may shed light on the origin of the X-ray flare emission (pure Synchrotron or SSC?)
and hence on the energetics in the immediate vicinity of the black hole candidate SgrA*.

Recent progress in detecting gravitational waves from merging black hole binaries  \citep{Abbott2016}
using LIGO\footnote{http://www.ligo.org/science.php}
shows that this window is also opening for exploring  the immediate vicinity of SgrA*.
The interaction of stars and stellar remnants amongst each other as well as with the large mass
at the center of the Milky Way may provide tools to explore the nature of SgrA*
using gravitational waves \citep[e.g.][]{Amaro-Seoane2012, Freitag2003, Pierro2001}.

\begin{table*}[!htbp]
\begin{tabular}{cllll}\hline \hline
label   & necessary condition                                                                & referring to  \\ 
        &                                                                                    & section \\ \hline
$N_1$   & Is object at nominal position of SgrA*?                                            & \ref{Distance}\\
$N_2$   & Is size of emitting region in SgrA* sufficiently small?                            & \ref{Size} \\
$N_3$   & Is mass of SgrA* in agreement with SMBH masses?                                    & \ref{Mass} \\
$N_4$   & Does the distance to SgrA*  place it at the center of the Milky Way?               & \ref{Distance} \\
$N_5$   & Is the manipulative success for SgrA* similar to other SMBH candidates?            & \ref{manipulative} \\
$N_6$   & Is a bright fast jet originating from SgrA*?                                       & \ref{jet} \\
$N_7$   & Do we detect a merger ringing signal in gravitational waves from SgrA*?            & \ref{ringing} \\
$N_8$   & Do we detect an exceptionally bright flare from SgrA*?                             & \ref{Variability} \\
$N_9$   & Do stars and pulsars close to SgrA* give indications for a SMBH?                                     & \ref{pulsars} \\
$N_{10}$& Is the spectrum of the surroundings of SgrA* what es expect from a SMBH?          & \ref{Spectrum} \\
$N_{11}$& Do we detect a photon ring in SgrA* in addition to orbiting matter?                & \ref{Horizon} \\
$N_{12}$& Do VLBI images of SgrA* show a shadow as expected for a SMBH?                     & \ref{shadow}, \ref{Horizon},  \ref{jet},  \ref{futureobs} \\
$N_{13}$& Do we detect photo-center motion of SgrA* with NIR- and/or mm-radio-interferometry?& \ref{futureobs} \\
$N_{14}$& Can we differentiate fo SgrA* between jet components and hot-spot?         & \ref{Spin}, \ref{Horizon}, \ref{futureobs} \\
\hline \hline
\end{tabular}
\caption{
Table of possible necessary conditions that can be combined to result in 
a sufficient condition required to call SgrA* a SMBH.
The necessary conditions have been formulated as logical entities for which we can attribute the 
locigal values ``true" or ``false" within the theoretical predictions for supermassive black holes
in section \ref{predictions}.
}
\label{tab:criteria}
\end{table*}

\section{Synthesis: Combining the Results} 
\label{synthesis} 

We can now proceed to apply the Eleatic Priciple to the investigation of SgrA* and then ask the question:
How good a case for being a SMBH is it and how good a case can it become?
Based on the preceding discussion we have described a number 
of critical observational results that are required as necessary conditions of calling
SgrA* a SMBH.
A priori it is not clear if all necessary conditions are known
such that their simultaneous fulfillment results in a
sufficient condition for the existence of a SMBH.
The conditions discussed in this article are listed in Tab.\ref{tab:criteria}.
Conditions N$_1$ to $N_4$ are rather fundamental, one might even say, ``technical" conditions 
that are required for most of the combinations to form sufficient conditions.
Here, basically only the necessary conditions $N_5$ to $N_8$ have the character of a ``smoking gun",
in the sense that a certain observational criterium of spectacular nature points at the existence 
(or rather ``action") of a supermassive black hole.
Conditions  $N_9$ to $N_{14}$ are observational results that are substantially more difficult to obtain.
They rely on either special instrumentation and/or the source SgrA* being ``cooperative", in the
sense that the required event occurs very rarely and/or the black hole environment must be very
clean to observe the necessary expected  or predicted phenomena.
Certainly, not all of these need to be or - given their probability of occurrence - can be 
combined and fulfilled to serve as a convincing sufficient condition for calling SgrA*
a SMBH.
To a certain extent this relieves us from the pressing question if  
all necessary conditions for the proof of existence are known and fulfilled
- as a variety of choices can be used to result in a 
sufficient condition for the existence for an  SMBH at the center of the Milky Way.

In combining the different necessary conditions listed in 
Tab.\ref{tab:criteria} we can first ask the question 
if there are any conditions that can be extracted from
specific combinations of sufficient conditions following equation~(\ref{eq:5}). 
Trivially, condition $N_1$ must always be
fulfilled, as the aim is making a statement on SgrA*.
However, as we will see, this appears to be the only general condition,
but the requirements on  the accuracy with which the light at different wavelengths 
can be associated with the position of the large mass varies strongly from case to case.
We also note that not all conditions listed in Tab.\ref{tab:criteria} are strong 
necessary conditions, e.g. the spectrum ($N_{10}$) of the surroundings of SgrA*  
may equally well be produced by other scenarii involving a very compact massive 
object that is not necessarily an  SMBH.
A schematic representation of the different combinations discussed in the 
following is shown on the left side of Fig.~\ref{fig0}. 

Possible combinations that may lead to convincing sufficient conditions appear to be:

$S_1 =  N_1 \wedge N_2  \wedge N_3  \wedge N_4  \wedge N_5  \wedge N_{10}$ : 
The {\bf ``manipulative success"} can only work in a statistical sense and is difficult
to be applied to SgrA* as an individual source. However, if solid observables 
for demonstrating ``manipulative success" can be found and SgrA* is included in the 
sample, then this could be looked upon as a sufficient condition.
Depending on these observables mass, the position and distance will be needed.
In addition to mass and distance, the overall spectrum and a small radio source 
size will be needed to support 
the presence of a SMBH in the framework of a statistical 
compilation centered around the ``manipulative success" approach.

$S_2 =  N_1  \wedge N_3  \wedge N_4  \wedge N_6$ : 
The {\bf detection of a jet} showing relativistic properties like superluminal motion 
of individual components is a clear indication of an
accretion phenomenon onto a very compact massive object.
It needs to be combined with the presence of a large mass to be indicative of a SMBH identification.
Obtaining the mass needs the position of SgrA* and its distance.
A strong and fast jet implies strong accretion which makes it unlikely for fragile 
objects like neutrino and boson stars to persist and not to collapse into a SMBH.

$S_3 =  N_1  \wedge N_8$ : 
The {\bf detection of an exceptionally bright flare} from SgrA*
that may compensate a major part of its current underluminosity of 
10$^{-9}$ times the Eddington luminosity and
would certainly indicate the presence of a strong accretion 
event onto a very compact object like a SMBH. 
Such very bright outbursts of SgrA* have apparently occurred in the past.
A bright X-ray outburst of SgrA* within the recent $\sim$400 years 
is being discussed as the possible reason for the X-ray fluorescence emission 
from massive molecular clouds that are surrounding the Galactic Center
\citep{Clavel2013,Terrier2010, Revnivtsev2004, Sunyaev1998}.
\cite{Witzel2012}  explain this ($\sim$10$^{39}$ erg/s) outburst 
via an extreme value of their near-infrared statistics without the need for an extraordinary event.
Based on the flux frequency plot for SgrA* shown in Fig.19 of \cite{Witzel2012}  
and the conditions discussed therein,
the maximum near infrared brightness expected at 2$\mu$m wavelength 
would be of the order of 3~Jy or about 8.5 magnitudes.
It would occur every 10 to 100 million years.

$S_4 =  N_1 \wedge N_2 \wedge N_4  \wedge N_{11}$: 
If the accretion process is clean enough, one can speculate 
that in addition to orbiting matter    
a {\bf secondary photon ring} can be detected towards SgrA*.
Imaging of a photon ring could be done as a quasi-stationary image taken with VLBI at short mm-wavelengths
resulting in assessing the small intrinsic source size with minimum influence by scattering effects. 
Modeling results using the 
distance to SgrA* will result in a mass and compactness and will allow us to state that a 
sufficient condition for the detection of an extremely high mass concentration like that provided by
a SMBH has been fulfilled.
However, \cite{Vincent2015}
show that it may not be possible to uniquely conclude that it is in fact a SMBH
rather than a boson star.
Differences between the images of the secondary photon ring expected for a black hole and a boson star 
require VLBI imaging with a high dynamic range and the source being in a low state of activity to
allow for sensitive imaging.

\begin{figure*}
\begin{center}
\includegraphics[width=12cm]{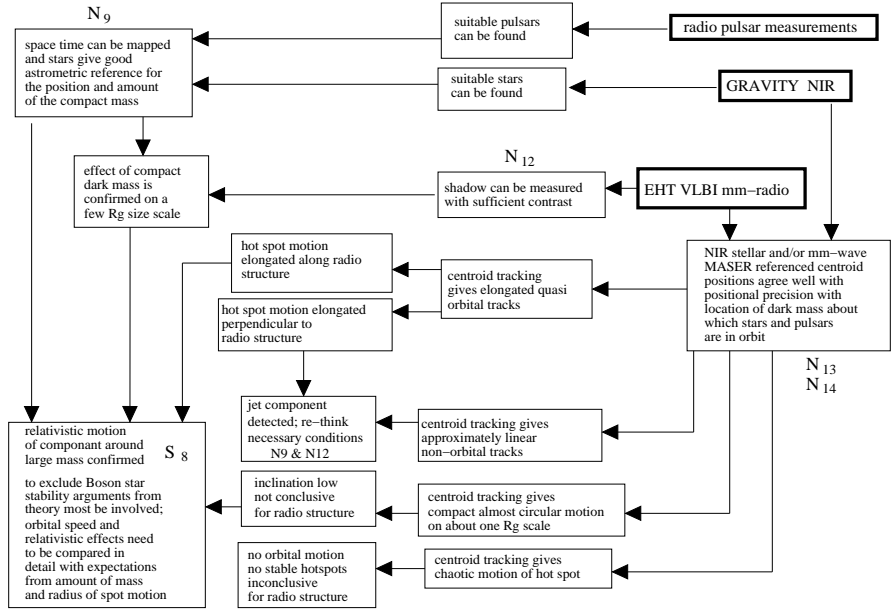}
\end{center}
\caption{Sketch of a decision flow chart for option $S_8$ to combine suitable 
necessary conditions.
\label{fig7}}
\end{figure*}

$S_5 =  N_1  \wedge N_7$ : 
The {\bf detection of a merger ringing signal} in gravitational waves  
is a clear indication for a highly relativistic phenomenon. The ringing frequency 
would allow us to pinpoint the evolved masses and size scales without previous 
knowledge of these quantities.
Ringing indicates a merger process and hence violent accretion which makes it 
unlikely for fragile objects like neutron or boson stars to remain stable (see above).
However, the small expected merger rates make the detection of a ringing 
signal very unlikely. 
In this context, it is the comparison to the 
observational results for other galactic nuclei which may help to 
understand the situation for SgrA*.

$S_6 =  N_1  \wedge N_9$ : 
If the {\bf search for stars and pulsars} towards SgrA* is successful, 
then the mapping of spacetime will tell the mass and size scales. 
The orbital parameters of stars and pulsars will allow us to
independently determine the distance and involved size scales.
Detailed mapping of spacetime with pulsars or stars within a
few 1000 Schwarzschild radii 
and measuring the quadrupole moment of SgrA* 
\citep[e.g.,][]{Psaltis2012,Psaltis2016}
also allows us to convincingly distinguish the black hole case 
from a boson star case.
If GRAVITY finds stars with small separations to SgrA*, this may be the only experiment 
that allows us to simultaneously indentify the location of the SgrA* 
continuum emission with the position of the central mass, 
i.e. the object about which the stars are in orbit.
This seems to be within reach even with the the currently observable 
stars S2, S02-1, and S38 \citep{Boehle2016, Parsa2017}.

\begin{figure}
\begin{center}
\includegraphics[width=12cm]{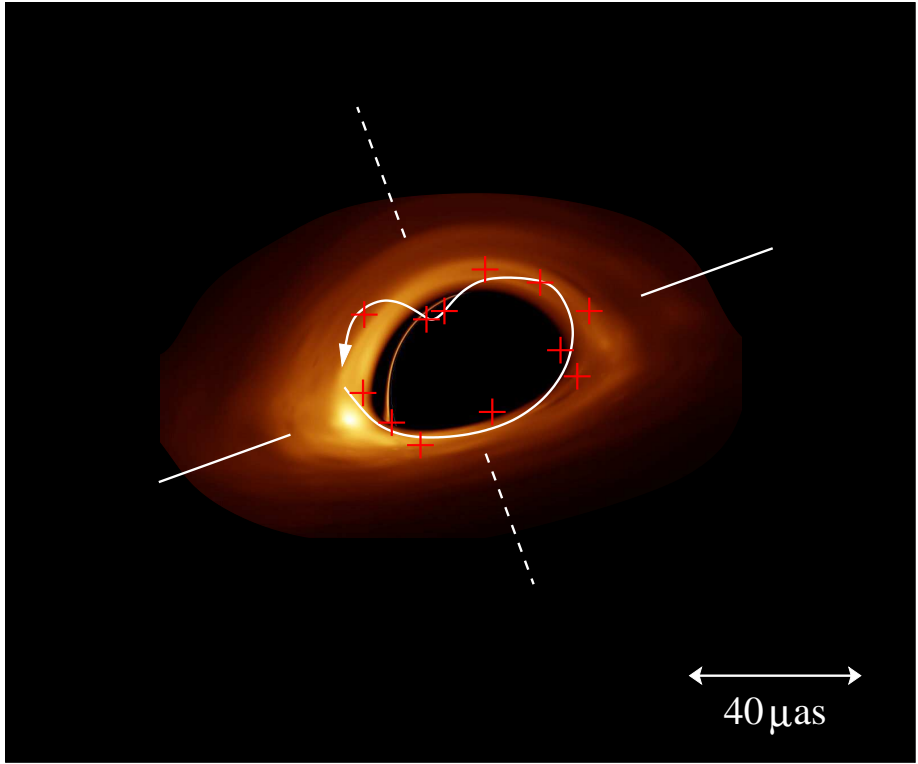}
\end{center}
\caption{
Photocenter motion compared to a disk model.
The example of a NIR photo-center motion as planned to be measured with the GRAVITY interferometer at the VLTI 
is taken from \cite{2008Paumard} and \cite{paumard2005}. The simulation describes  
the apparent trajectory of flare events assuming material orbiting a non-rotation black hole
at an inclination of 45$^o$ on the last stable orbit at a distance of 3~$R_S$ from the center.
Lensing (including multiple images), relativistic beaming and Doppler effect are included 
in the relative positioning of the resulting data points (red crosses) following the orbital track 
\citep[white line; further details in ][]{2008Paumard}.
The image \citep{bursa2007} is assumed to represent a mm-VLBI data disk model that shows 
luminous material for radii beyond the last stable orbit.
The dashed and straight white arrows indicate the directions perpendicular and along the radio structure
that we refer to in the text.
\label{fig6}}
\end{figure}

\begin{figure}
\begin{center}
\includegraphics[width=12cm]{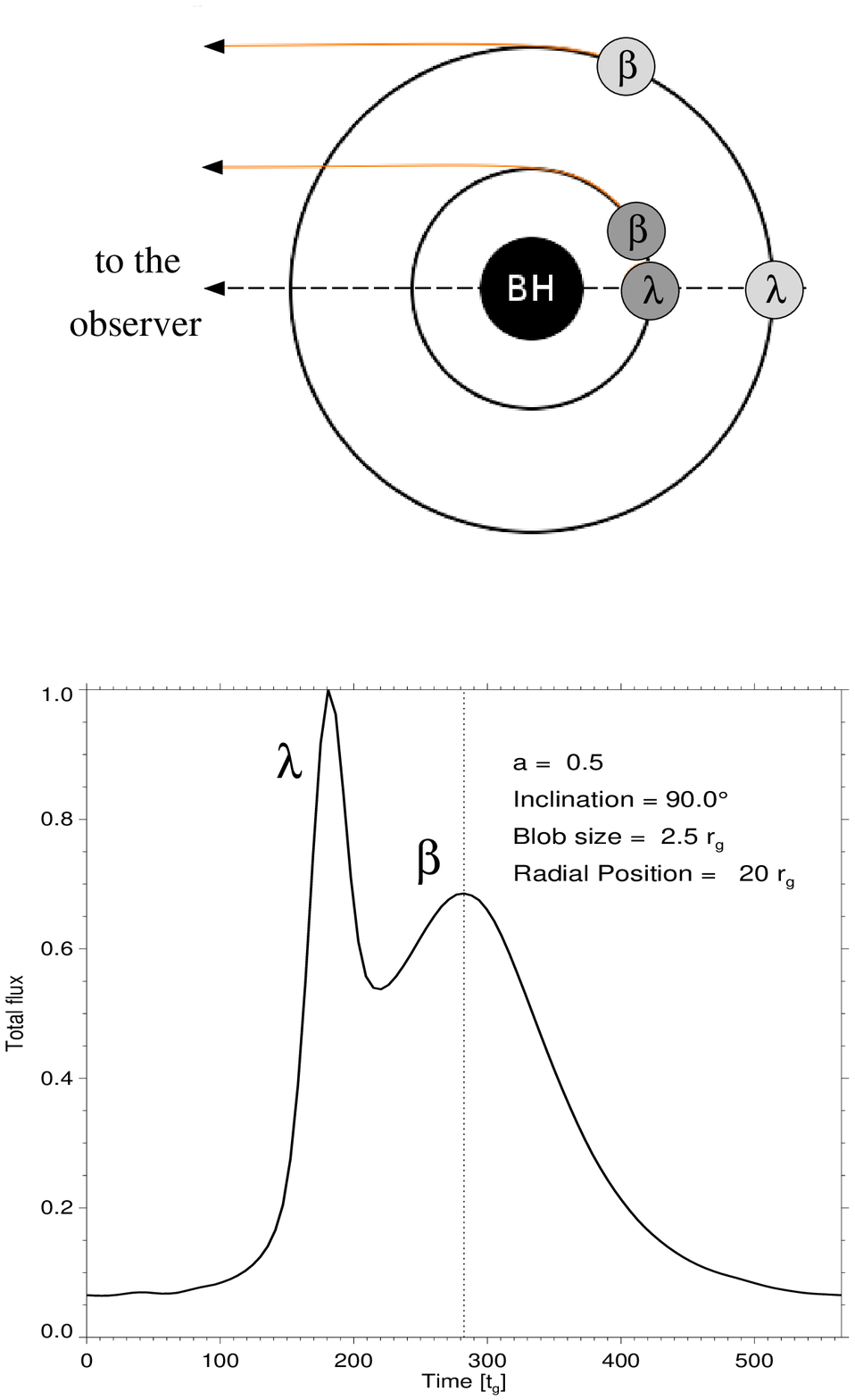}
\end{center}
\caption{
At the top we show the geometrical arrangement of position at which beaming ($\beta$) and lensing ($\lambda$)
may lead to detectable subflare structures. These are marked for an example case in the light curve on the bottom.
The light curve is plotted for a full orbit of the plot as a function of gravitational
time units in $t_g = r_g / c$. For a black hole mass of 4.3$\pm$0.3$\times$10$^6$\solm the unit 
$t_g$ is of the order of 20 seconds.
\label{fig6.2}}
\end{figure}

$S_7 =  N_1 \wedge N_2 \wedge N_{12} \wedge N_{13}  \wedge N_{14}$:
{\bf Combining mm-VLBI and}:
{\bf NIR-inter- ferometry results}:
mm–VLBI images will be very compact and croissant shaped both for orbiting
spots and for jet components \citep{Dexter2012a, Dexter2012b},
because the formation of a sensitive VLBI image takes at least several hours
and this is largely in excess of the orbiting time scale of spots
close to the ISCO (less than about half an hour) in the case of SgrA*. Hence, VLBI will deliver at best a
quasi-stationary image of the shadow which could be resulting from a jet
dominated image as well \citep{Dexter2012b, Dexter2012a}.
It is not clear, how well low contrast spots orbiting the central mass can be tracked
with a high time resolution using VLBI. A limited and variable uv-plain coverage as well as 
radiative transport phenomena may make it difficult to reliably produce densely sampled time series of 
snapshot images.
Also, \cite{Vincent2015} show that shadow-like structures can be produced by 
other extremely high mass concentrations like e.g. boson stars.
SgrA* is variable at all wavelengths indicating variable accretion and possibly structural changes 
in the continuum light emitting central region.
Hence, it is unclear under what conditions small differences between the expected
shadow and accretion disk structure between a black hole and a boson star scenario can be detected with VLBI imaging.

{\it The NIR GRAVITY experiment} will not be able to spatially resolve the 
immediate vicinity (1-2~mas; see below) region around SgrA* (i.e. its accretion wind or a temporary disk/jet structure).
The resolution may not be sufficient to determine if the emission is
coming from the ISCO or from a jet nozzle that may be significantly offset from
the black hole and hence may not reflect the situation in its immediate vicinity
\citep[see the cases of M87 and SgrA* discussed by][]{Nakamura2013, Falke2009}.
In the radio regime such an apparent offset depends on the position at which the
jet becomes optically thin. At shorter wavelengths, however, investigations of
stellar and SMBHs in the framework of ``lamp post" models 
indicate that the bright base of the jet may be located about 2 Schwarzschild radii (or more) above 
the accretion disk, i.e. at some distance from the disk's ISCO orbit
\citep{Miller2015,Emmanoulopoulos2015}.
The radio and NIR position of SgrA* agree to within 10~mas
\citep{Reid2003}, i.e. about 1000 Schwarzschild radii.
The inferred position agrees with the position of the mass concentration
about which the stars orbit with projected separations from SgrA* of about 1~mas i.e. 
about 100 Schwarzschild radii \citep{Reid2003, Ghez2008, gillessen2009a}.
Hence, there is room for a significant offset  between
a possible jet nozzle from the ISCO.

However, a great advantage of the NIR interferometric measurements is that it 
will allow us to track and time resolve the centroid motion of SgrA*
with integration times of a few minutes.
Following the motion of a hot plasma blob in orbit around (or in
transversal motion towards or away from) SgrA* requires being able
to distinguish between bright spots separated only a few Schwarzschild radii
from each other. This implies an accuracy of a few ten of microarcseconds.
Hence, VLBI images alone $(N_1  \wedge N_{12})$ 
as well as  NIR centroid measurements by themselves $(N_1  \wedge N_{13})$ 
may not allow for differentiation between a foot-point of a precessing jet 
and an orbiting hot-spot ($N_{14}$).
However, the NIR result will allow us to discover an oscillating 
bright spot motion orientated along the VLBI structure 
(see Fig.~\ref{fig6}). This is expected for inclined orbital motion of 
spots at a distance of a few Schwarzschild radii from the black hole
(compare Fig.~\ref{fig4} left and Fig.~\ref{fig5} right as well as 
\cite{2010Zamaninasab} and \cite{Vincent2014}).
One may expect periodic motion of spots at larger radii perpendicular 
to the VLBI structure in the case of a foot-point of a precessing jet
(see Fig.~\ref{fig6} and \ref{fig4}).
The quality of the alignment of the two disk representations 
in Fig.~\ref{fig6} will depend on the quality of the
infrared and radio reference frame. However, perfect alignment (i.e. superposition of the 
black hole mass locations) is not required in order to distinguish between the different directions of motions.
For inclinations larger than about 45$^o$ and orbits close to the last 
stable orbit the radii of the photo-center motion 
become more circular and are less well suited for comparison with 
the radio structure.
For inclination of less than 45$^o$ the consecutive flare peaks from boosting and lensing may be observable.
In Fig.~\ref{fig6.2} we show the geometrical arrangement for positions at which these light
amplifications occur. In addition, we show the lightcurve for an example case for which the time difference between the
two sub-flare events will be of the order of 30 minutes.
For the example in Fig.~\ref{fig6.2} the corresponding positional ``jump" between these two events amounts to about 4 R$_g$, i.e. 
about 400$\mu$as, hence well detectable with the 10-100$\mu$as astrometric precision expected for GRAVITY.
For smaller orbits the amplification points move closer together.
The presence of a clearly defined blob, i.e. hot-spot, as well as moderately low inclinations are required
in order to observe this effect.
The advantage is that the live time of the blob only needs to be of the order of a fraction of an orbital period.
In the decision flow shown in Fig.~\ref{fig7} this case would correspond to centroid tracking
of (short) elongated quasi orbital tracks which would then be perpendicular to the spin axis.
Therefore, a combination of both, mm-VLBI and NIR-interferometry with GRAVITY,
will lead to a strong sufficient condition to demonstrate
the existence of a heavy mass on the scale of one Schwarzschild diameter,
i.e. a black hole.

{\it Stellar MASER emission observed with mm-VLBI} 
will also allow us to conduct centroiding experiments. 
Since the mm-emission is optically thick, those  observations
will be truly complementary to the centroid experiments on the optically thin NIR emission.
43~GHz and 86~GHz masers have been found in the stellar atmospheres
of a number of late type stars in the central stellar cluster in the immediate vicinity of SgrA*
(\cite{Reid2003,reid2003b} and Borkar et al. 2016 in prep.).
\cite{Doeleman1998} find that 86~GHz masers mirror the 43~GHz positions.
SiO masers originate for Mira type sources 2-4 stellar radii i.e. 4-8 AU from the star. 
The maser spot shell will then be distributed over less than 1 mas at the distance of SgrA*.
The 200-800~km baselines allow for an angular resolution of up to $\sim$1mas at 43~GHz and 86~GHz
and allow us to phase reference (or in beam reference) SgrA* with respect to the masers. 
Hence, phase referenced observations can be used to search for and track
orbiting spots as well as low-surface brightness features like jets or out- and inflows.

$S_8 =  N_1 \wedge N_2 \wedge N_9  \wedge N_{12} \wedge N_{13}  \wedge N_{14}$:
\cite{Psaltis2016} show that the {\bf stellar and pulsar results combined
with those of the shadow of the black hole} result in a
high accuracy test for SgrA* being a SMBH.
Accepting that the VLBI images do not easily allow us to track a motion 
on the orbiting time scale at the ISCO, the combination of 
the pulsar, black hole shadow and the NIR- and mm-interferometry result appear as the
most unique combination to convincingly show that a SMBH exists.
Measuring the detailed influences of the Kerr metric on the orbits of pulsars 
close to SgrA*, a boson star as an alternative explanation could also be excluded 
\citep{psaltis2015a,Psaltis2016}.
In Fig.~\ref{fig7} we show a 
sketch of a decision flow chart for option $S_8$ to combine suitable necessary conditions.
Details of the chart are described in option $S_7$ as well.

\begin{figure*}
\begin{center}
\includegraphics[width=12cm]{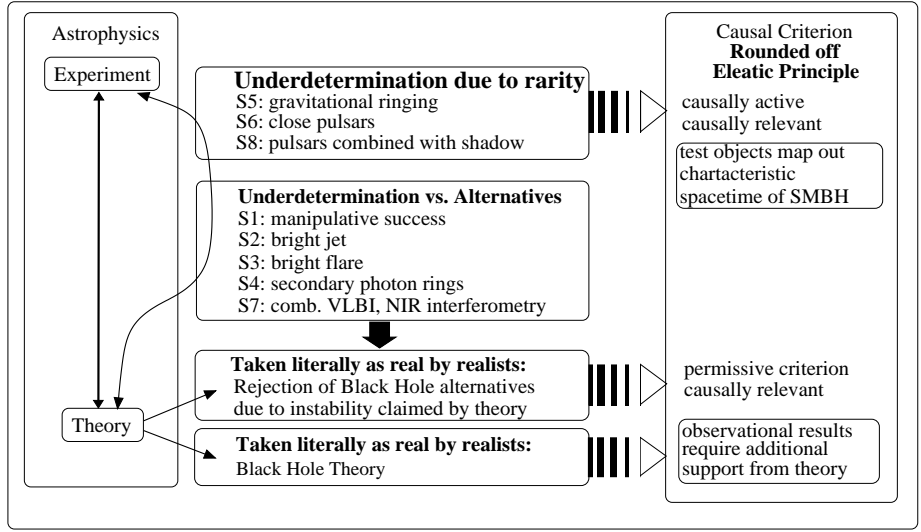}
\end{center}
\caption{Linkage between experiment and theory interpreted via the concept of
realism, underdetermination and a ``rounded out" version of the Eleatic Principle,
here shown with respect to the results of our investigation.
For comparison see also Fig.\ref{figm00} which we adopted here for the case of the 
Galactic Center SMBH.
\label{fig8}} 
\end{figure*}

\section{Summary and Conclusion} 
\label{conclusion} 

It appears that acceptance and proof of the presence of a SMBH is in a state quite comparable to e.g. the acceptance of
the existence of atoms and molecules at the beginning of the past century.
The Eleatic Principle as a heuristic causal criterion to probe the existence of an entity 
has been very helpful in that process.
A detailed analysis of how the necessary conditions need to be combined 
to a convincing case
will also be essential in the case of the SMBHs.

{\it A challenge for astrophysics can be summarized in the question:} 
How clean are the observational cases that may serve as
logical entities for the causal criterion test?
How rare are the ``smoking gun" cases and the coincidence of fully operational, complex 
experiments and a ``cooperative" source, such that the expected or predicted phenomena 
do occur and can actually be measured.
The ``manipulative success" approach ($S_1$) may be contaminated by statistical noise as 
one needs a finite number of  objects of which SgrA* will only be one.
The success of the method will certainly also depend on the choice of the 
observables on which statistics is done.
Occasional bright jets or flares ($S_2$ and $S_3$) are probably very rare,
but are likely to have occurred during the past of SgrA* (see estimates and references above).
In these cases the cleanness of the observables will depend on the physical properties
of the jet (e.g. its orientation to the line of sight) or the brightness and duration of the flare.
Similarly gravitational wave ringing ($S_5$) characteristic for interactions of large masses 
may be rare or very difficult to measure.
The detection of a secondary photon ring  ($S_4$) will only be possible if the accretion 
process is well ordered and not too violent such that high contrast observations 
of the vicinity of the SMBH can be carried out successfully.

{\it A conceptual or philosophical challenge} 
arises from the usage of a two stage approach based on the 
concept of
"(Anti)Realism and Underdetermination" and a modified form of the Eleatic Principle
(section \ref{subsection:antirealism}, \ref{App:section:bridge}, \ref{App:section:EleaticPrinciple}).
This approach is mainly driven by the 
theoretical and observational necessity that forces us to deal with the observational 
consequences imposed by theory (e.g. the fact that it takes an infinite amount of time to reach
the event horizon) and the deficiencies in the measuring process. There is hope that
the  latter can be minimized by instrumental and observational progress and an elegant and fruitful
combination of different methods to approach the observational problems.
The application of these concepts then may give us enough confidence to accept the validity or
reality of the black hole theory - either literally or based on the observational results 
that point at the reality of (in our case) supermassive black holes.
Trust in theory and observational results may then lead us to involve 
causation as a clue to the existence of supermassive blck holes 
in the way it is laid out in section
\label{subsection:bridge} and Appendix~\ref{App:section:EleaticPrinciple}
and discussed in detail in, e.g., \cite{Colyvan1998} and \cite{Marcus2015}.
With respect to the results of our investigation we have depicted 
the situation in Fig.\ref{fig8}.
As shown in this figure, our results are successfully mapped out by this 
combination of concepts.

Underdetermination arises both from the rarity of the expected event and from the fact that
the observations alone can probably not clearly discriminate the black hole case against possible alternatives.
Here, only the theoretical finding can help, that alternatives do not lead to stable solution.
As in the case of a standpoint that a pure realist may take concerning black hole theory, such a 
reasoning then requires a permissive criterion as proposed by 
\cite{Sellars1963} and  \cite{Psillos2011b}.
This leads to applying the ``rounded out" version of the Eleatic Principle 
described by Colyvan in 1998 \citep{Colyvan1998}.
However, even though in the end one may need additional support through 
theoretical concepts, it needs to be stressed that here we are not concerned
with the question of the ontological state of the theoretical concept but rather 
with maximizing the acceptance of an identification as a SMBH 
through the comparison to the theoretical concept of it
to test the properties of an observable entity against a theoretical concept
in order to identify it in a commonly agreable way with the object
described by this very theoretical concept
(see paragraph ``Theory taken as being real" in section \ref{App:section:bridge}).
A further challenge arises from the complexity of the 
observables and the corresponding quality of the necessary and 
sufficient conditions\footnote{
In the context of the SMBH candidate SgrA* at the
Galactic Center and the large number of achievable and potential observables,
the conceptual strategy laid out in section~\ref{PhilConcepts}
and depicted in Fig.~\ref{fig0} is not trivial.
If several combinations of necessary conditions may lead to 
a convincing statement concerning the existence of an  SMBH
at the center of the Milky Way,
then a conceptual or philosophical challenge 
arises through the fact that some are more convincing than others.
A mere measurement of a relativistic jet might be less valued than 
the detection of an pulsar extremely close to the black hole.
In this case, we would have to refrain from Boolean logic, in which 
only true and false (1 or 0) are allowed as values of the necessary and 
sufficient conditions $N$ and $S$ and the resulting 
value $K$ (see section \ref{formal}).
A many-valued logic in which the values of these conditions 
can be any real number between 0 and 1 is then a possible way
and equations~(\ref{eq:1}) to (\ref{eq:6}) must be modified accordingly.

Using the notation from section \ref{formal} equation~(\ref{eq:4})
would then for the sufficient condition $\lambda'$ translate into:
\begin{equation}
K_{\lambda'}=\Pi_{i=1}^{\mu(\lambda')}~~~N_{\kappa_{i,\mu(\lambda')}}
\end{equation}
The value of K will be high in case that a combination of necessary conditions have been met well.
Then the favorite sufficient condition would have a value of:
\begin{equation}
K=max\left[K_{\lambda} | \lambda= 1 , \nu \right] 
 =max\left[\Pi_{i=1}^{\mu(\lambda)}~~~N_{\kappa_{i,\mu(\lambda)}} | \lambda= 1 , \nu \right]~~. 
\end{equation}
}.

Several of the sufficient conditions we outlined in section~\ref{synthesis} 
are more biased towards the mere detection of a very high mass concentration than
the distinction between an  SMBH and one or several of the discussed alternative explanations.
With evermore complex conceptual models for compact masses that may serve as an alternative 
to a SMBH it becomes also more difficult to judge
if one falls into some of the pit holes presented on the right panel in 
Fig.~\ref{fig0}: 
Do we oversee necessary conditions ($\eta$)?; 
Are we combining the rights ones (e.g. $\epsilon$ or $\eta$)?; etc.

High chances of success for delivering convincing results concerning the 
achievement of technical success and the repeatability of the experiments are represented by the
combination of millimeter and infrared interferometry in possible combination with 
the pulsar measurements ($S_6$, $S_7$ and $S_8$).
If GRAVITY finds stars closer in - or even with the current 
availability of known high velocity S-stars - GRAVITY may be the only experiment 
that allows us to determine the location of the SgrA* emission 
with respect to the central mass position, i.e. the object about which the stars are in 
orbit. A (temporarily) bright accretion disk may be too turbulent and too much effected by 
optical depth and light bending effects. Hence the usefulness of determining the location of the
mass center may be too much model dependent (compared to the determination of a stellar orbit).
Images of the material close to the SgrA* supermassive black hole will be shortly obtained
by the EHT.

If detailed mapping of the spacetime is not possible due to the 
lack of suitable pulsars or stars,
it appears that the only argument that allows us to easily reject the
boson star (or other alternative models) possibility is the large number of 
additional requirements (e.g. the existence of special particles) that need to be fulfilled to
sustain its existence,  in particular the fact that it is very 
fragile and - if stability conditions are disturbed - it will collapse into a 
black hole.

Hence, it appears that SgrA* is indeed already an outstanding case to prove 
the existence of SMBHs. 
Further support for the existence of supermassive black holes in general may come 
from gravitational wave experiments like eLISA.
However, due to its proximity, SgrA* highest angular resolution observations translate into
high linear resolution at the location of SgrA*. This results in a large number of 
necessary conditions that can be probed and fulfilled (see Tab.\ref{tab:criteria}).
New instruments will allow for new high precession measurements in the very near future that
are certain to improve the quality of SgrA* as a convincing showcase
for a SMBH that can be ``accepted as real". 
However, it appears that the experiments or observational effects that may come up with the most convincing 
arguments for the existence of supermassive black holes are also the ones that 
are the most difficult to achieve or the rares to happen.
Yet, despite all the great efforts put forward by researchers, it is necessary that 
SgrA* and its immediate surroundings provide us with useful flares 
as well as stars and pulsars with suitable orbits for
maximizing the usefulness of the measurements.

\section*{Acknowledgments}
We thank 
Sybille Anderl (IPAG Grenoble) for valuable comments and support on the philosophy of science sections,
Georgi Dvali (LMU Munich) for constructive and valuable discussions and input,
and 
Grischa Karssen (University of Cologne) for contributing Fig.\ref{fig6.2} and part of the corresponding discussion.
This work was supported in part by the Deutsche
Forschungsgemeinschaft (DFG) via the Cologne Bonn Graduate
School (BCGS), the Max Planck Society through the
International Max Planck Research School (IMPRS) for Astronomy
and Astrophysics. 
Part of this
work was supported by fruitful discussions with members of
the European Union funded COST Action MP0905: Black
Holes in a Violent Universe and the Czech Science Foundation
DFG collaboration (No.~ 14-37086G) and with members
of the European Union Seventh Framework Program
(FP7/2007-2013) under grant agreement No.~312789, Strong
Gravity: Probing Strong Gravity by Black Holes Across the
Range of Masses.

\newpage

\appendix

\section{An example for establishing existence claims}
\label{App:subPhilQuestions}

As an example for establishing existence claims in the (recent) history of physics,
we look at the case of atoms and molecules.
The first case we want to look at - because it has been widely discussed in philosophy 
of science - is the final acceptance of the existence of molecules/atoms due to the work 
of Jean Perrin on Brownian motion in the early 20th century
as the collision with the quick atoms or molecules in the gas or liquid.
Brownian motion itself was first discovered by the Scottish botanist Robert Brown in 1827
as random motion of particles.

Perrin received the Nobel Prize for physics in 1926 for his work 
that ``put a definite end to the long struggle regarding the real existence of molecules"
(C.W. Oseen, member of the Nobel Committee for Physics).
So, how was Perrin able to establish the existence of molecules/atoms and what can we learn 
from this for the question of the existence of black holes? 
Atoms and molecules among other places played a major role in the kinetic theory of heat. 
However, there were debates about whether atoms/molecules were merely a useful fiction that 
yielded (in many cases) the right result or whether they should be accepted as real. 
For some time it appeared to be the case that this question could not be resolved by 
experimental means. In the early 20th century (partly through the work of Einstein) 
atoms and molecules became accessible, i.e. new theoretical means were devised so as to bring 
the hypothesis into the reach of established experimental methods 
\citep[see][ ``Making Contact with Molecules: On Perrin and Achinstein" ]{Psillos2011}.

In making the question of atoms/molecules not only experimentally accessible but to decide 
it positively, three features proved to be important:
\begin{itemize}
\item[i.]
The assumption that atoms/molecules exist puts constraints on observational data: 
The assumptions could only be true if some observable magnitudes had certain very definite values 
(Avogadro's number/constant).
This appears to be an essential point in a number of cases. What is asked for here are novel predictions, 
i.e. the prediction of phenomena, which would be very unlikely if atoms/molecules did not exist 
(compare the prediction of the deflection of light by the sun as evidence for  GR;  
Karl Popper tried to capture this by his notion of a high degree of falsifiability). 
The prediction has to be risky.

\item[ii.]
On rival accounts, i.e. on the denial of atoms/molecules, the value of Avogadro's number would 
have been completely different.
Hence, the value of Avogadro's constant is dependent on the existence of atoms.

\item[iii.]
The relevant value was experimentally established in a number of different/independent ways.
Avogadro's constant is tested in several different ways and today there are about 60 different 
and independent ways to measure it \citep{Bethege2012},
making use of e.g. electrical charge and current, in determinating the properties of gas, liquids and crystals.
\end{itemize}

Following the ground breaking work of Perrin, today the acceptance of atoms is supported by
direct imaging of molecules, atoms or even their bounds \citep[e.g.][]{Oteyza2015, Iwata2015}
\footnote{The case of pulsars and the case of the identification of sources 
in which gamma-ray bursts originate may be a further example in which establishing existence 
claims is or has been important in the the recent history of physics.}.

\section{From (Anti)realism and Underdetermination to a Causal Principle}
\label{App:section:bridge}

How for theory and observations the concepts of ``Realism" and ``Underdetermination" may be 
linked to causation is outlined in the following:

{\it Theory taken literally as being true:}
Black holes are extremely simple objects and hence, ideally suited to be regarded as 
a kind of theoretical concept that historically has the tendency to be 
taken literally as real by realists
(see below and statements by Stathis Psillos and Wilfrid Sellars 
\citep{Psillos2011b,Sellars1963}).
The problem is that 
other theories could be regarded as being similarly simple 
(although they are often more complex as they require the definition of additional
quantities like particle masses and interaction forces etc.).
The fact that one cannot easily distinguish between them experimentally, leaves enough 
freedom to being unable to decide if some other entities implied by a literal reading of the theory, 
e.g. a compact object like a black hole and its possible alternatives, are indeed real 
(e.g. the shadow of such a compact object, i.e. the presence, depth and size of an 
emission free or sparse region close to the position of the compact object; see section \ref{shadow}).

Here, also the possibility of an experimental comparison between theory and implied entities 
(if interpreted as predicted phenomena like speeds of stars or emitting blobs, 
timescales associated with consecutive events, or structures like the shadow)
in a way relieves many philosophers 
(especially if facts or statements are being taken literally)
from the ``commitment to a host of 
entities with a (to say the least) questionable ontic status: numbers, geometrical points, 
theoretical ideals, models and suchlike" \citep{Psillos2011b}.

Therefore, we must clarify that at this point we are not concerned
with the question of whether there is reality to the theoretical concept of a black hole as such,
for instance as a concept within the theory or relativity.
The goal of the observational and experimental astrophysics is
to test the properties of an observable entity against a theoretical concept
in order to identify it in a commonly agreeable way with the object
described by this very theoretical concept.

{\it Reality of entities derived from theory:}
The entities implied by the theory of black holes 
(here taken as observational results or as implied physical scenarii, i.e.
how stars orbit the black hole or - in the near future - how exactly the black hole shadow looks like) 
can also be regarded as real, as they all come (are derived) indeed 
"from theory and ... there is no theory-free standpoint from which"
 \citep{Psillos2011b}, these entities can be viewed.
This means, as valuable observational predictions are usually all derived from theory there is no theory-independent 
criterion of reality that can be used in an accepted way.
The goal is to define and successfully observed entities i.e. predictions, for which 
a black hole plays an indispensable role in their explanation. 

While this is clearly the preferred situation for the physicist (i.e. a successful comparison 
between observational results or the properties of implied physical scenarii) this
is also true for theory as such from the viewing angle of realists that historically 
tend to read (take) theories literally as real.
As this can be explained out of theory it can be
regarded as a permissive criterion \citep{Sellars1963}
that particularly does not disallow abstract entities from being real.
Stathis Psillos \citep{Psillos2011b}
points out that ``this explanatory criterion should not be confused with a causal
criterion", e.g. in the form of the Eleatic Principle (see below), ``according to which
being causally active, that is having causal powers, is a criterion of objecthood.
Causal efficacy may well be a mark of reality, but not everything 
that is real is causally efficacious" \citep{Psillos2011b}.
The former would be good for experimentalists, the latter would be supported by black hole theory.
Here, Psillos points at entities that are causally idle but causally relevant.
Hence, he describes Colyvan's ``rounded out" version of the Eleatic Principle 
(Colyvan 1998 \citep{Colyvan1998} and 
the following section \ref{App:section:EleaticPrinciple}).

{\it Causation as a clue to probe existence:}
For observed entities, may they be derived from theory or not,
causation may be taken as a clue to make their existence plausible beyond doubt.
David M. Amstrong states in his reply to Reinhardt Grossmann 
in the correspondence on the ``Ontology and the Physical Universe"
\citep{CumpaTegtmeier2009}
that if one does not ``postulate an entity [...]
which is not required in ones account of causation", then
indeed causation can be used as a partial clue to the existence.
The word ``partial" is used here as it needs to be seen 
how the analysis of the causation needs to be done in every particular case.
Goal of the observational astrophysics is to optimize this 
aspect.

As Andrew Newman states:
"Talking about existing and talking about being real are
just different ways of talking about the same thing"
\cite{NewmannAndrew2002}.
As supporting examples he quotes Gottlob Frege and Bertrand Russell 
who are guided to things that are real by 
syntax \citep[e.g. in ][]{Carnap1968},
W.V. Quine, who demands that something must be quantifiable 
if it is called real \citep[e.g. in ][]{DeccockHorsten2000},
and D.M. Amstrong demanding causal significance for real 
things \citep[e.g. in ][]{NewmannAndrew2002}.
Causal significance or relevance is an essence of any form of the 
Eleatic Principle (see below).

Andrew Newman points out \citep{NewmannAndrew2002}:
"Alex Oliver claims that the Eleatic Principle is ambiguous because
there is an epistemological reading of 'reason' and a metaphysical reading of
'reason', mostly worried about the existence of causally inactive entities."
This criticism might appear to be applicable to aspects of the 
black hole physics, however, all experiments aim at consolidating
causal activity.

{\it Reality and the Eleatic Principle:}
In the case of black holes, we may make use of the possible claim that the physical
world is causally closed, i.e. that all genuine causes of physical events are
physical causes \citep[e.g.][]{Montero2003}.
Hence, in the case of black holes we do not 
run into the problem that this particular science entity is not reducible to
a physical entity.
Even if the reduction to such an entity is currently not fully conducted, all 
instrumental and observational efforts aim at improving the performance of that reduction.

\section{The Eleatic Principle}
\label{App:section:EleaticPrinciple}

If underdetermination can be fought or even partially overcome,
then causation may be used to further underline 
the realism or existence of an entity in a generally acceptable way.
This involves the usage of a causal criterion that may be in the form of the Eleatic Principle 
\citep[for a general overview see e.g.][]{Colyvan1998,Colyvan2001}.
\cite[][]{Colyvan1998} gives a consice definition of the classical Eleatic Principle
"An entity is to be counted as real if and only if it is capable 
of participating in causal processes".

The principle is named after a Greek school in lower Italy Elea
('$E \lambda \acute{\epsilon} \alpha$)
closely linked to the philosophers Parmenides, Zeno and
Xenophanes of Colophon (going back to a figure in Plato's dialogue Sophistes).
As a philosophical conceptual aspect, the School of Elea rejects any
epistemological criteria simply based on sensual experiences (Fig.\ref{fig00}).
According to the  Eleatic Principle, we should be realists 
about whatever manifests itself in virtue of having effects 
(in `A World of States of Affairs' by David M. Amstrong \citep{Armstrong2001}, see also the discussion in 
"Quining Naturalism" by  Huw Price \citep{Price2007}).
The use of the Eleatic Principle in our context needs some clarification.
The classical version of the Eleatic Principle uses and 
connects (in a more time-like fashion) mainly causally active entities.
In this version non-causal entities like abstract mathematical 
and physical concepts are regarded as causally idle and therefore
they are not considered as suitable entities.
This problem has been explained in detail by Mark Colyvan in 
his article ``Can the Eleatic Principle be Justified?"
\citep{Colyvan1998}.
He investigates several justification attempts and shows 
how and when they fail or what their drawbacks are.
In order to avoid these problems he suggests to use a ``rounded out"
version of the Eleatic Principle in which also causally idle 
entities can be used.
This ``rounded out" version contains the classical 
version of the Eleatic Principle (Fig.\ref{fig00}).
In the case of the Galactic Center and supermassive black holes in general, 
involvement of at least the ``rounded out" version of the
Eleatic Principle is justified  \citep[see below and][]{Colyvan1998}.

Mark Colyvan's version of the ``rounded out" Eleatic Principle does not stand alone.
He points out that this more general principle that he proposes 
has great similarities to Quine's thesis that ``we are only ontologically 
committed to all and only the entities that are indispensable to our current best scientific theories"
 \citep[see][]{Quine1948}.
Indispensable entities then include both causally idle and causally efficacious entities.
A critical discussion of the Eleatic Principle with a focus on the
role of mathematical objects and Colyvan's defense of Quine's indispensability argument
is given by \cite{Marcus2015}.
It should also be mentioned that the entities we connect in the 
synthesis section~\ref{synthesis} 
have by themselves a rather complex nature.
In most cases they are the result of 
a logical combination of causally idle and active entities
and have a structure similar to that shown in equation~(\ref{eq:5}).
Causally idle (but ally relevant) entities occur e.g. through the 
theory of image formation in all wavelength domains or in relativistic concepts that are
used to extract expected situations and compare them to measurements.
In particular the ``data inversion" problem in the image formation theory of interferometry might 
well serve as an example of true ``underdetermination" in astrophysics.
Causally efficacious entities occur e.g. 
in the form of stars, pulsars, emitted radiation
or astrophysical instruments and telescopes,
all of which can be located in spacetime in contrast to the causally idle entities.

\section{Time scales and luminosities of gravitational wave ``ringing"}
\label{App:ringing}

The analysis is done based on the formalism presented in 
\cite{MisnerThorneWheeler1973} and \cite{ShapiroTeukolsky1983}.
The tidal disruption radius $r_{\rm{T}}$ for main-sequence stars with mass $m_{\star}$ and radius $R_{\star}$ is larger than the Schwarzschild radius $R_{\rm{S}}$ of the supermassive black hole,
\begin{equation}
  r_{\rm{T}}\approx 19 R_{\rm{S}}\left(\frac{R_{\star}}{R_{\odot}}\right)\left(\frac{M_{\bullet}}{4\times 10^6\,M_{\odot}}\right)^{1/3}\left(\frac{m_{star}}{1\,M_{\odot}}\right)^{-1/3}\,,
\end{equation}
so we cannot expect any gravitational wave event from main-sequence star inspirals, since they will be tidally disrupted. On the other hand, Schwarzschild radius is larger than tidal disruption radius for white dwarfs with $R_{\star}=0.01\,R_{\odot}$, neutron stars with $R_{\star}=10^{-5}\,R_{\odot}$, and stellar black holes, so these are not subject to tidal disruption.

A circularized orbit of a compact star with the semimajor axis $a_{\star}$ around the supermassive black hole $M_{\bullet}$ emits gravitational waves at frequency,
\begin{equation}
  \nu_{\rm{GW}}=\frac{2}{P_{\rm{orb}}}=0.18\,\rm{mHz}\,\left(\frac{M_{\bullet}}{4\times 10^6\,M_{\odot}}\right)^{1/2}\left(\frac{a_{\star}}{10\,\rm{R_{\rm{S}}}}\right)^{-3/2}\,,
\end{equation}  
reaching $\sim 1.1\,\rm{mHz}$ at innermost stable circular orbit.

The characteristic scaling amplitude for the EMRI event at the distance of the Galactic center is

\begin{equation}
 h_0=4\times 10^{-18}\,\left(\frac{M_{\bullet}}{4\times 10^6\,M_{\odot}}\right)^{2/3} \left(\frac{m}{1\,M_{\odot}}\right) \left(\frac{D}{8\,\rm{kpc}} \right)^{-1} \left(\frac{\nu_{\rm{GW}}}{1\,\rm{mHz}} \right)^{2/3}\,,
\end{equation}

and the effective metric perturbation is given by the amplitude $h_0$ times the square root of the number of
cycles the EMRI spends in band with a bandwidth $\Delta \nu$ centered at the frequency $\nu$. The number of cycles $N$ is given by $N=\nu^2/\dot{\nu}=\nu \tau$, where the characteristic evolution time-scale $\tau \equiv \nu/\dot{\nu} = 8/3(t_{\rm coal}-t) \propto \mathcal{M}^{-5/3} \nu^{-8/3}$ and $t_{\rm coal}$ is the coalescence time. The `chirp mass' may be expressed as $\mathcal{M}=(M_{\rm \bullet}m_{\star})^{3/5}/(M_{\bullet}+m_{\star})^{1/5} \approx M_{\bullet}^{2/5} m_{\star}^{3/5}$, where the approximation holds for the EMRI. Finally, the metric perturbation or basically the signal is proportional to $S \propto h_0 (\mathcal{M} \nu)^{-5/6}=h_0 M_{\bullet}^{-1/3} m_{\star}^{-1/2} \nu^{-5/6}$.

The expected luminosity $L_{\rm GW}$ averaged over one period and evaluated at the innermost 
stable circular orbit \citep{PetersMathews1963} is

\begin{equation}
  L_{\rm{GW}}=1.9\times 10^{43}\,\rm{erg\,s^{-1}}\left(\frac{m_{\star}}{1\,M_{\odot}}\right)^2 \left(\frac{M_{\bullet}}{4\times 10^6\,M_{\odot}}\right)^3 \left(\frac{a_{\star}}{3\,R_{\rm{S}}}\right)^{-5}\,,
  \label{eq_gw_luminosity}     
\end{equation} 
which is valid for circular orbits. In general, the equation \ref{eq_gw_luminosity} depends strongly on the orbital eccentricity  \citep{PetersMathews1963}. The incoming flux $F_{\rm GW}$ may be expressed as a function of the amplitude $h_0$ and the observed frequency $\nu_{\rm{GW}}$,

\begin{equation}
  F_{\rm GW}\simeq 0.005 \left(\frac{\nu_{\rm GW}}{1\,{\rm mHz}}\right)^2 \left(\frac{h_{0}}{4\times 10^{-18}\,} \right)^2\,{\rm erg\,cm^{-2}\,s^{-1}}\,.
  \label{eq_flux}
\end{equation}

The coalescence (merger) time-scale $\tau_{\rm{merge}}$ depends on eccentricity of the orbit $e$. For a circular orbit, 

\begin{equation}
\tau_0(a_0) \approx 7800\,\rm{yr}\left(\frac{M_{\bullet}}{4\times 10^6\,M_{\odot}}\right)^{-2} \left(\frac{m_{\star}}{1\,M_{\odot}}\right)^{-1} \left(\frac{a_0}{10\,R_{\rm{S}}}\right)^4\,,
\end{equation}
where $a_0$ is an initial semi-major axis. For initially highly eccentric orbits, $e\approx 0.99$, it reduces to $\tau_{\rm{merge}}=768/425 \tau_0(a_0)(1-e_0)^{7/2}\approx 0.016\,\rm{yr}$.

 The detection of an EMRI event associated with the Galactic center black hole would enable to precisely measure the mass and the spin of the black hole with an independent measurement. The waveform could also reveal deviations from the Kerr metric as well as a possible different character of a central compact object (e.g. a boson star). An exemplary waveform for an EMRI event for the ratio $m_{\star}/M_{\bullet}=10^{-4}$ and the black hole spin $J=0.998$ is depicted in Fig.~\ref{fig_emri}. The inspiral starts at four gravitational radii.

The EMRI events from the Galactic center will be within the detection sensitivity limits of the planned LISA and eLISA space-base interferometers that will be capable to detect the gravitational wave events with low frequencies in the range $\nu_{\rm{GW}}=0.1\,\rm{mHz}$--$1\,\rm{mHz}$. Although the likelihood to detect an EMRI event for the Galactic center is rather low during the mission lifetime, \cite{2004CQGra..21S1595G} estimate that LISA will be able to detect $\sim 2$ EMRI events of $1.4\,M_{\odot}$ compact objects (white dwarfs and neutron stars) per cubic Gpc per year for black holes with similar masses as Sgr~A*.

\newpage
\bibliography{philo}
\bibliographystyle{spbasic}      % basic style, author-year citations
\end{document}